\documentclass[12pt,preprint]{aastex61}

\shorttitle{Coronal Loops}
\shortauthors{van Ballegooijen et al.}

\begin{document}

\title{The Heating of Solar Coronal Loops by Alfv\'{e}n Wave Turbulence}

\author[0000-0002-5622-3540]{A. A. van Ballegooijen}
\affiliation{5001 Riverwood Avenue, Sarasota, FL 34231, USA}
\author{M. Asgari-Targhi}
\affiliation{Harvard-Smithsonian Center for Astrophysics, 
60 Garden Street, Cambridge, MA 02138, USA}
\author{A. Voss}
\affiliation{School of Computer Science, University of St~Andrews, Jack Cole Building, North Haugh, St~Andrews, Fife, KY16 9SX, Scotland, UK}
\affiliation{Harvard-Smithsonian Center for Astrophysics, 
60 Garden Street, Cambridge, MA 02138, USA}

\begin{abstract}
In this paper we further develop a model for the heating of coronal loops by Alfv\'{e}n wave turbulence (AWT). 
The Alfv\'{e}n waves are assumed to be launched from a collection of kilogauss flux tubes in the photosphere at the two ends of the loop. Using a three-dimensional magneto-hydrodynamic (MHD) model for an active-region loop, we investigate how the waves from neighboring flux tubes interact in the chromosphere and corona. For a particular combination of model parameters we find that AWT can produce enough heat to maintain a peak temperature of about 2.5~MK, somewhat lower than the temperatures of 3 -- 4 MK observed in the cores of active regions. The heating rates vary strongly in space and time, but the simulated heating events have durations less than 1 minute and are unlikely to reproduce the observed broad Differential Emission Measure distributions of active regions. The simulated spectral line non-thermal widths are predicted to be about 27 $\rm km ~ s^{-1}$, which is high compared to the observed values. Therefore, the present AWT model does not satisfy the observational constraints. An alternative ``magnetic braiding" model is considered in which the coronal field lines are subject to slow random footpoint motions, but we find that such long period motions produce much less heating than the shorter period waves launched within the flux tubes. We discuss several possibilities for resolving the problem of producing sufficiently hot loops in active regions.
\end{abstract}


\section{INTRODUCTION}

The nature of coronal heating is one of the unsolved problems in solar physics
\citep[see reviews by][]{Zirker1993, Schrijver2000, Aschwanden2005, Klimchuk2006,
DeMoortel2015}. The energy for coronal heating is believed to originate in the convection
zone below the photosphere, but the details of how this energy is transported to and dissipated
in the corona are not well understood. One possibility for heating coronal loops is that the
sub-surface convective flows cause twisting and braiding of coronal field lines, which leads
to the formation of thin current sheets in the corona where magnetic reconnection can take
place \citep[e.g.][]{Parker1972, Parker1983, Berger1991, Berger1993, Priest2002,
Janse2010, Berger2015, Wilmot-Smith2015, Pontin2015, Pontin2017}. The reconnection likely
proceeds in a burst-like manner, producing ``nanoflares" \citep[][]{Parker1988}, and the corona
may be heated by the combined effect of a large number of such nanoflares occurring at
different times and positions within the coronal loop \citep[e.g.,][]{Cargill1994, Klimchuk2010,
Cargill2015}. The strongest heating occurs in active regions, which have broad temperature
distributions with temperatures in the range 1 to 6 MK \citep[e.g.,][]{Kano1996, Winebarger2011,
Warren2011, Warren2012, DelZanna2015b}. In the nanoflare model these broad temperature
distributions are explained by assuming that an observed coronal loop consist of multiple
threads, each in a different state of its temperature evolution \citep[][]{Cargill2004,
Cargill1997, Patsourakos2009, Cargill2015, LopezFuentes2015}. Coronal radio bursts may
be a signature of nanoflare heating \citep[][]{Mercier1997}. Ohmic dissipation of braided magnetic
fields is assumed to be the main cause of coronal heating in MHD models of coronal loops
\citep[e.g.,][]{Rappazzo2007, Rappazzo2008, Dahlburg2016} and entire active regions
\citep[e.g.,][]{Gudiksen2005, Bingert2011, Bourdin2015}.

In the photosphere outside sunspots the magnetic field is highly intermittent and is concentrated
in small magnetic flux elements (``flux tubes"), which have kilogauss field strengths and widths of
a few 100~km or less \citep[e.g.,][]{Stenflo1973, Solanki1993, Berger1996, Berger2001}. These
flux tubes may be formed by convective collapse \citep[][]{Parker1978, Spruit1979, Nagata2008,
Fischer2009}. In plage regions the magnetic flux concentrations have typical field strengths of
about 1500 G, and the flux tubes expand with height \citep[][]{MartinezPillet1997, Berger2004,
Buehler2015}. Strong downflows occur in the immediate surroundings of these flux tubes.
Neighboring flux tubes merge at some height in the upper photosphere or
low chromosphere \citep[e.g.,][]{Bruls1995}. Transverse MHD waves can be
produced in such flux tubes as a result of their interactions with granule-scale convective flows
\citep[e.g.,][]{Matsumoto2010, Mumford2015}. Such waves may propagate upward into the solar
atmosphere and dissipate their energy in the corona \citep[][]{Alfven1947, Wentzel1974,
Hollweg1978, Parnell2012, Arregui2015, Cranmer2015}. Alfv\'{e}n waves are of particular
interest because they can propagate over large distances in the corona before giving up their
energy. Transverse waves have been observed in the corona above the solar limb
\citep[][]{Tomczyk2007, Tomczyk2009, Threlfall2013, Morton2015}, in the swaying motions
of spicules \citep[][]{DePontieu2007}, in network jets on the solar disk \citep[][]{Tian2011,
Tian2014}, and in the solar wind \citep[][]{Coleman1968, Belcher1971, Matthaeus1990,
Bale2005, Borovsky2012}. The observed amplitudes of Alfv\'{e}n waves at the coronal base
are sufficient to heat and accelerate the solar wind \citep[][]{DePontieu2007, McIntosh2011},
and Alfv\'{e}n waves are believed to be the main driver of the fast wind \citep[e.g.,][]
{Suzuki2005, Cranmer2007, Verdini2007, Chandran2011}.
Alfv\'{e}n waves may also be responsible for heating coronal loops \citep[e.g.,][hereafter
paper I]{Moriyasu2004, Antolin2010, vanB2011}. These models assume that the waves
originate in the photosphere, not as a by-product of nanoflares in the corona. Therefore,
the connection to the photosphere is very important in wave heating models.

There are many observational constraints on coronal heating models. First and foremost,
the model should reproduce the observed coronal temperatures and densities \citep[e.g.,][]
{Winebarger2011, Warren2011, Warren2012}, as well as the spatial and temporal variations
of these quantities \citep[e.g.,][]{Viall2011, Viall2012}. Many coronal loops seen in EUV and
X-ray images have nearly constant cross-sections \citep[e.g.,][]{Klimchuk2000,
LopezFuentes2006, LopezFuentes2008}, which appears to be in conflict with the basic idea
that the magnetic field lines expand with height in the corona. Observations of the ``moss" at
the ends of hot loops provide constraints on the energy losses by downward conduction
\citep[][]{Fletcher1999, Martens2000, Warren2008, Winebarger2011}, and measurements
of plasma density can provide constraints on the spatial distribution of the heating
\citep[][]{Fludra2017}. Second, the model should
reproduce the observed spectral line widths, which are broadened in excess of their thermal
widths \citep[e.g.,][]{Doschek2007, Young2007, Tripathi2009, Warren2011, Tripathi2011,
Tian2011, Tian2012a, Tian2012b, Doschek2012, Brooks2016, Testa2016}. The observed
non-thermal broadening provides constraints on unresolved reconnection outflows in the
corona, and on the velocity amplitudes of MHD waves. Third, the model should be consistent
with observations of the braiding or twisting of the coronal field lines, or the lack thereof
\citep[e.g.,][]{Schrijver1999, Brooks2013}. Finally, the model should explain observations
of coronal ``rain" \citep[e.g.,][] {Antolin2012, Antolin2015}, which provide information on the
heating and cooling of the coronal plasma.

Another, often overlooked constraint comes from observations of the motions
of magnetic flux elements in the photosphere \citep[e.g.,][]{Muller1994, Schrijver1996,
Berger1996, Berger2001, Berger1998, vanB1998, MansoSainz2011, Wedemeyer2012}.
Such horizontal motions are believed to be responsible for the braiding of coronal magnetic
field lines \citep[][]{Parker1972, Parker1983} and/or the generation of transverse MHD
waves \citep[][]{Vigeesh2012, Mumford2015}.
Photospheric ``bright points" are often used as proxies for kilogauss flux elements
\citep[e.g.,][]{Berger1996, Nisenson2003, Abramenko2011, Utz2010}.
\citet[][]{Chitta2012} used high-cadence observations of isolated bright points and measured
an rms velocity $v_{\rm rms} = 1$ $\rm km ~ s^{-1}$ and a velocity autocorrelation time
$\tau_{\rm c} = 30$~s, but this represents only the motions on very short time scales.
On longer time scales the random motions are often characterized as ``random walk" with
a certain photospheric diffusion constant $D$. In magnetic network and plage regions 
$D \approx 60$ $\rm km^2 ~ s^{-1}$ on a time scale of a few thousand seconds \citep[][]
{Berger1998}, and on a time scale of several days $D$ is in the range 100 -- 250
$\rm km^2 ~ s^{-1}$ \citep[][]{DeVore1985, Wang1988, Schrijver1990, Komm1995,
Schrijver1996, Hagenaar1999}. These values of $D$ put severe constraints on the rate at
which magnetic energy can be injected into the corona by random footpoints motions
\citep[][and paper~I]{vanB1986}.

In this paper we focus on the wave heating model, but we also compare it with the magnetic
braiding model. In paper~I only a single magnetic flux tube was
considered, so the interactions between neighboring flux tubes were ignored. In contrast,
in magnetic braiding models the different flux tubes are assumed to be wrapped around each
other by photospheric convective flows, and thin current sheets are expected to develop at the
interfaces between the flux tubes \citep[e.g.,][]{Parker1983, Wilmot-Smith2009, Rappazzo2013,
Pontin2015, Ritchie2016}. It is desirable to include the effects of multiple flux tubes also in the
wave heating models. Here we extend the model of paper~I to include the interactions between
neighboring flux tubes. We investigate how the transverse waves from neighboring flux tubes
interact in the chromosphere and corona. Current sheets can develop at the boundaries between
the tubes because the tangential component of velocity is not expected to be continuous across such
boundaries for waves originating in different flux tubes. The dissipation of these boundary currents
may increase the wave heating compared to models with a single flux tube.
We also make further improvements to the treatment of the chromosphere-corona transition
regions (TRs) at the two ends of the coronal loop. In paper~I the TRs were treated as
discontinuities in plasma density, and their effect on the Alfv\'{e}n waves was described in terms
of reflection and transmission coefficients. In the present work the TRs are fully resolved along
the loop.

The paper is organized as follows. Previous work on the wave heating model is summarized in
section 2. In section 3 we describe a new wave heating model for coronal loops in which the
magnetic field of the loop is anchored in a collection of kilogauss flux tubes in the photosphere.
The model describes the dynamics of Alfv\'{e}n waves in such a complex magnetic structure.
In section 4 simulation results are presented for one particular set of model parameters. We
study the spatial distribution of the heating, the amplitude of heating events at various positions
along the loop, and the effects of the waves on spectral line profiles. In section 5 we consider
a magnetic braiding model in which the photospheric flux tubes are omitted and the footpoint
motions occur on longer time scales, but with exactly the same background atmosphere as the
first model. Therefore, the heating rates predicted by the two models can be directly compared.
We find that for the same velocity of footpoint motions (about 1 $\rm km ~ s^{-1}$) the magnetic
braiding model produces less heating than the wave heating model. The results are further
discussed in section 6.

\section{WAVE HEATING MODELS}
\label{HEATING}

Transverse MHD waves can be dissipated in a variety of ways. In the presence of density
variations across the magnetic field lines, the waves can be damped by phase mixing and
resonant absorption, which cause wave energy to be transferred to smaller scales
\citep[][]{Heyvaerts1983, Poedts1990, DeGroof2002, Goossens2011, Goossens2012,
Goossens2013, Pascoe2011, Pascoe2012}. Torsional Alfv\'{e}n waves can interact
nonlinearly with the background plasma to produce parallel flows and shocks, which
provide another dissipation channel \citep[][]{Kudoh1999, Moriyasu2004, Suzuki2005,
Antolin2010}. Counter-propagating Alfv\'{e}n waves are subject to nonlinear interactions
\citep[][]{Iroshnikov1963, Kraichnan1965}, which produce turbulent cascades \citep[e.g.,][]
{Shebalin1983, Goldreich1995, Maron2001, Cho2002}. The resulting Alfv\'{e}n wave
turbulence (AWT) is believed to play an important role in the heating of the corona in both
open and closed magnetic structures \citep[][]{Matthaeus1999, Oughton2001, Dmitruk2003}.

A key feature of AWT in a low-beta plasma is that it is highly anisotropic, with velocity
perturbations nearly perpendicular to the background magnetic field, and with
perpendicular length scales much smaller than the parallel ones. Therefore, AWT in the
solar corona is expected to be quite different from isotropic turbulence in ordinary fluids.
Strong AWT can develop even when the velocity amplitude $\delta v_\perp$ of the
counter-propagating waves is much smaller than the Alfv\'{e}n speed ($\delta v_\perp \ll v_A$).
Reflection-driven AWT is believed to be responsible for producing the fast solar wind
\citep[e.g.,][]{Cranmer2007, Verdini2007, Chandran2011}. Detailed 3D MHD simulations of
turbulent waves in the acceleration region of the solar wind have been described by
\citet[][]{Perez2013} and \citet[][hereafter paper~V]{vanB2016, vanB2017}.
Density variations across the magnetic field are known to exist both in coronal loops and in
the solar wind. Most turbulence modeling is based on the reduced MHD approximation
\citep[][]{Strauss1976, Strauss1997}, which produces a passive cascade of density and
magnetic-field-strength fluctuations at scales larger than the ion-gyro radius
\citep[e.g.,][]{Schekochihin2009}. However, most solar models developed so far neglect the
density fluctuations altogether \citep[e.g.,][papers~I and IV]{Perez2013}. Therefore, the
turbulence models do not yet include the important effects of
phase mixing and resonant absorption. Conversely, most studies of resonant absorption use
linearized versions of the MHD equations, and therefore neglect the effect of the waves on
the background density variations, as well as the nonlinear couplings between
counter-propagating transverse waves. Therefore, a comprehensive description of the physical
processes that cause wave energy to be transferred to smaller scales is not yet available.

Alfv\'{e}n wave turbulence may also be responsible for the heating of the chromosphere and
corona in active regions (paper~I). Alfv\'{e}n- and kink waves can be
produced in photospheric flux tubes as a result of their interactions with granule-scale convective
flows \citep[e.g.,][]{Mumford2015}. In paper~I we treated the flux tubes as having rigid walls, so
the transverse waves could be approximated as Alfv\'{e}n waves and simulated using the
reduced MHD approximation. We found that the waves propagate upward along the
expanding flux tube and reflect due to variations in Alfv\'{e}n speed with height; this led to the
development of AWT in both the chromospheric and coronal parts of the loop. In the corona the
counter-propagating waves are launched from both ends of the loop, so they have roughly
equal amplitudes and their nonlinear interactions are quite strong. It was found that the loops
typically observed in active regions can be explained in terms of AWT, provided the small-scale
footpoint motions in the photosphere have velocities of 1 -- 2 $\rm km ~ s^{-1}$ and time scales
of 60 -- 200 s.

Magnetic braiding does occur in the AWT model, but the braids are highly dynamic and not
close to a force-free state. This is a consequence of the fact that the lower atmosphere is
included in the model. The high density of the photosphere compared to the corona implies that
all perturbations tend to be wave-like, and quasi-static evolution is found only when the lower
atmosphere is omitted from the modeling \citep[][]{vanB2014}. Another key feature of the AWT
model is that the energy injected into the corona is dissipated on a time scale comparable to the
coronal Alfv\'{e}n travel time, which is 20 -- 60 s for typical active-region loops. In contrast, in the
nanoflare model it is assumed that magnetic free energy can be built up in the corona for
thousands of seconds before part of the energy is released in a reconnection event. This
requires that the current layers in the corona maintain a finite thickness, so that reconnection
does not occur prematurely and energy can build up over a longer period of time \citep[e.g.,][]
{Wilmot-Smith2009, Pontin2015, Ritchie2016}. Whether such energy build-up occurs or not
depends on the nature of the magnetic structures and flows in the lower atmosphere where
convective driving takes place. Therefore, it is important to include the lower atmosphere in
models for coronal heating.

The AWT model was further developed by constructing loop models for active regions
observed with the {\it Solar Dynamics Observatory} \citep[][hereafter papers II, III and IV]
{Asgari2012, Asgari2013, Asgari2014}. Papers II and IV used nonlinear force-free field
(NLFFF) modeling to obtain the magnetic field strength $B_0 (s)$ as a function of position $s$
along the loops, which is a key parameter in any heating model; paper III used potential-field
modeling. A set of field lines was
selected, and for each one the wave turbulence in a thin flux tube surrounding the selected field
line was simulated. It was found that the wave heating rate $Q(s,t)$ averaged over the loop
cross-section depends on the position along the loop, and varies with time in a burst-like manner.
The mean heating rate $\overline{Q}$ averaged over the coronal volume and over time was
about $10^{-3}$ $\rm erg ~ cm^{-3} ~ s^{-1}$, but the peak rates during heating events was about
one order of magnitude larger. The peak temperature $T_{\rm max}$ in the different loops
was predicted to be in the range 2.1 - 2.9 MK, but for any given loop the model predicts
only small temperature variations, $\Delta T \sim 0.1$ MK. This is due to the modest amplitude
of the heating events, and the fact that they are localized along the loop, so that
their effect is quickly diminished by thermal conduction. Therefore, in its present form the AWT
model cannot explain the higher temperatures of 4 - 6 MK observed in many active regions.
The higher temperatures can be obtained only by increasing the footpoint velocities to about
5 -- 6 $\rm km ~ s^{-1}$ \citep[][]{Asgari2015}, which seems beyond what may be expected for
magnetic footpoint motions in the photosphere on small spatial scales.

Paper~IV used the Extreme-ultraviolet Imaging Spectrometer (EIS) on {\it Hinode} to derive
observational constraints on Alfv\'{e}n wave amplitudes at the loop tops for an active region
observed on 2012 September 7. Spectral lines of Fe~XII, Fe~XIII, Fe~XV and Fe~XVI
were used to derive non-thermal velocities from the observed line widths. The authors found
wave amplitudes in the range 20 -- 34 $\rm km ~ s^{-1}$ for the Fe~XII 192 {\AA} line,
consistent with predictions from the AWT model. However, we now realize that the instrumental
line width may have been underestimated, so the non-thermal velocities may have been
overestimated. More complete EIS observations were presented by \citet[][]{Brooks2016},
who measured non-thermal line widths in 15 non-flaring active regions and found a mean value
of $17.6 \pm 5.3$ $\rm km ~ s^{-1}$.
Also, \citet{Testa2016} used the Interface Region Imaging Spectrometer (IRIS) and found
modest non-thermal velocities with an average of about 24 $\rm km ~ s^{-1}$ and a peak
of the velocity distribution at 15 $\rm km ~ s^{-1}$. \citet[][]{Hara1999} used a coronagraph
and spectrometer at the Norikura Solar Observatory to place constraints on the amplitudes of
Alfv\'{e}n waves in an active region observed above the solar limb \citep[also see][]
{Ichimoto1995}. They found that the non-thermal velocities for Fe~X 6374~{\AA},
Fe~XIV 5303~{\AA} and Ca~XV 5694~{\AA} are in the ranges 14 -- 20, 10 -- 18, and
16 -- 26 $\rm km ~ s^{-1}$, respectively. These observations of non-thermal velocity put severe
constraints on the AWT model for coronal heating.

In the above-mentioned works only a single magnetic flux tube was considered, corresponding
to one kilogauss flux element in the photosphere (magnetic flux $\Phi = 4.4 \times 10^{17}$ Mx).
Therefore, we implicitly assumed that a photospheric flux tube at one end of the coronal loop is
connected to a single flux tube at the other end. In reality the photospheric flux elements at the
two ends of a loop are uncorrelated and do not perfectly match up. In the next section we develop
a somewhat more realistic model containing multiple flux tubes that are not co-aligned at the two
ends.

\section{CORONAL LOOP MODEL CONTAINING MULTIPLE FLUX TUBES}
\label{multi}

\subsection{Lower Atmosphere}

\begin{figure}
\centering
\includegraphics[width=5in]{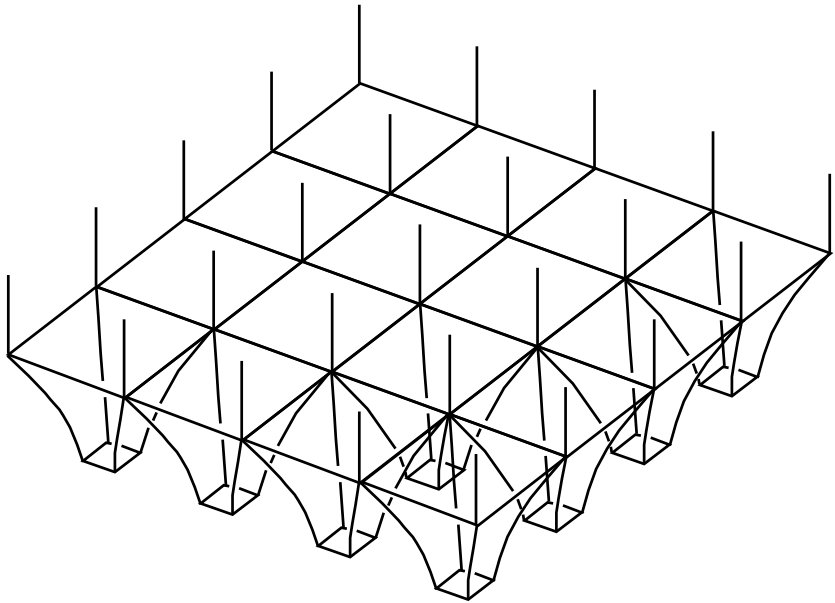}
\caption{Array of $4 \times 4$ flux tubes merging into a space-filling field at the positive polarity
end of a coronal loop. For clarity only 8 of the 16 flux tubes are drawn.}
\label{fig1}
\end{figure}

Coronal loops in active regions are anchored in the photosphere. The magnetic field in the
photosphere is highly intermittent and consists of a collection of kilogauss flux concentrations
separated by areas with much weaker fields \citep[e.g.,][]{Buehler2015}. In this paper we use
a very simplified model for the photospheric flux tubes, in which the tubes have square
cross-sections and are located on a square lattice, as illustrated in Figure~\ref{fig1}. An array
of $4 \times 4$ flux tubes is considered. Figure~\ref{fig1} shows the flux tubes at the positive
polarity end of the loop, and a similar array of flux tubes is present at the other end. The small
squares indicate the cross-sections of the flux tubes at the base of the photosphere, and the
large squares indicate the boundaries between flux tubes at the ``merging height" where
neighboring flux tubes come together. Just above the merging height the magnetic field is
approximately uniform, as indicated by the vertical lines.

In plage regions the photospheric flux tubes are densely packed, but spatially separated from
each other. In our model we assume that the medium between the flux tubes is field-free.  Let
$\overline{B_z}$ be the average magnetic flux density in the photosphere, i.e., the vertical
component of magnetic
field averaged over the flux tubes and the surrounding field-free medium. The magnetic field
emerges from the Sun at one end of the coronal loop, where $\overline{B_z} > 0$, and reenters
the Sun at the other end, where $\overline{B_z} < 0$. In this paper we construct a loop model
in which the magnetic field strength varies along the loop, and the average field strength in the
corona is 60~G (see next subsection). In this model the average magnetic flux density at the
base of the photosphere is $| \overline{B_z} | = 181.6$~G, which is typical for the magnetic flux
densities found in plage regions \citep[e.g.,][] {Buehler2015}. The flux density at the merging
height is slightly lower because of the overall expansion of the loop.

Each flux tube has a magnetic field strength $B_0 (z)$ that decreases with height $z$ in the
photosphere. The height $z = 0$ is defined as the {\it average} height of the surface where
the optical depth $\tau = 1$ at  wavelength $\lambda = 5000$~{\AA} in the continuum.
\citet[][]{Buehler2015} observed magnetic flux concentrations in a plage region, and found field
strengths of about 1520 G at optical depth $\log (\tau) = -0.9$. In kilogauss flux elements the
surfaces of constant $\tau$ are depressed relative to the surrounding photosphere (Wilson
depression). Therefore, in the present model we assume a field strength $B_0 (0) = 1500$~G
at the base of the photosphere ($z = 0$). The magnetic filling factor at height $z$ is approximately
given by $f(z) \approx | \overline{B_z} | / B_0 (z)$; in the present model $f(0) = 0.12$ at the base
of the photosphere. The filling factor increases with height, and reaches unity at the so-called
merging height ($z = z_m$), which lies in the upper photosphere or low chromosphere
\citep[e.g.,][]{Bruls1995}. Above this height the different flux tubes come together to fill the
available volume. Such merging of flux tubes into a space-filling field occurs at both ends of
the coronal loop. Therefore, in our model the merged field extends from the chromosphere
at one end of the loop to the chromosphere at the other end.

The flux tubes expand with height because of the stratification of the photosphere. To describe
this expansion, we use the thin-tube approximation \citep[][]{Spruit1976, Defouw1976}.
The flux tubes are assumed to be in pressure balance with their surroundings:
\begin{equation}
\frac{B_0^2}{8 \pi} = p_{\rm ext} (z) - p_{\rm int} (z) ,  \label{eq:pmag}
\end{equation}
where $p_{\rm int} (z)$ is the internal gas pressure, and $p_{\rm ext} (z)$ is the pressure in
the field-free external medium. The atmospheres outside and inside the flux tubes are assumed
to be in hydrostatic equilibrium:
\begin{eqnarray}
p_{\rm ext} (z) & = & p_{\rm ext,0} \exp \left[ - \int_0^z \frac{dz^\prime}
{H_p (z^\prime)} \right] ,  \label{eq:pext} \\
p_{\rm int} (z) & = & p_{\rm int,0} \exp \left[ - \int_0^z \frac{dz^\prime}
{H_p (z^\prime)} \right] ,  \label{eq:pint}
\end{eqnarray}
where $H_p (z)$ is the pressure scale height in the photosphere, and $p_{\rm ext,0}$
and $p_{\rm int,0}$ are the external and internal gas pressures at the base of the photosphere
(the ``external" medium is present only below the merging height). Here we neglect
temperature differences between the flux tubes and the external medium. At the base of
the photosphere the external pressure $p_{\rm ext,0} = 1.274 \times 10^5$ $\rm dyne ~ cm^{-2}$
(Spruit 1977), and the internal pressure $p_{\rm int,0} =  0.379 \times 10^5$ $\rm dyne ~ cm^{-2}$.
We use simple analytic models for the temperature and mean molecular weight as functions
of height in the solar atmosphere, which allows us to compute the pressure scale height
$H_p (z)$ and magnetic field strength $B_0 (z)$.

We then determine the merging height $z_m$ as the point where $B_0 (z)$ equals the field
strength $B(s)$ from the global loop model described in the next subsection. We find that the
merging height is located at $z_m \approx 520$ km in the temperature minimum region between
the photosphere and chromosphere. At the merging height $B_0 (z_m) \approx 176.6$ G,
slightly lower than the flux density $\overline{B_z}$ at the base of the photosphere because
of the overall expansion of the loop with height.
We assume that at the base of the photosphere the flux tubes have widths $w_0 = 400$ km, 
which is consistent with the observed sizes of magnetic flux concentrations in plage regions
\citep[e.g.,][]{Buehler2015}.  The magnetic flux of each tube is $\Phi = w_0^2 B_0 (0) =
2.4 \times 10^{18}$ Mx. The width $w_t (z)$ of each flux tube increases with height to
conserve magnetic flux, $w_t (z) = w_0 \sqrt{ B_0 (0) / B_0 (z) }$. At the merging height
$w_t (z_m)$ = 1,165 km, consistent with the typical distances between flux concentrations
in plage regions \citep[e.g.,][]{Buehler2015}. In our model $4 \times 4$ flux tubes come together
at the merging height, so the width of the merged field is $w_m (z_m) = 4 w_t (z_m) \approx$
4,660 km.

The merged field contains many separatrix surfaces corresponding to the boundaries
between flux tubes at the merging height (the large squares in Figure~\ref{fig1}). In our model
these are true separatrix surfaces, not quasi-separatrix layers, because in the photosphere
the flux tubes are separated by field-free regions, so they are topologically distinct. There are
two sets of separatrix surfaces, one from each end of the coronal loop. We assume that the
flux-tube structures at the two ends are shifted relative to each other by half the width of a flux
tube, $w_t (z_m)/2$, in both the $x$ and $y$ directions. Hence, the separatrix surfaces from
one end of the loop are located inside the flux tubes at the other end, and each flux tube is split
into 4 distinct flux systems. Therefore, the magnetic structure considered here contains 64
different flux systems (four times the number of flux tubes). In the presence of transverse
waves, electric currents can develop at the boundaries between these flux systems.

\subsection{Large-Scale Structure of the Merged Field}
\label{global}

On a larger scale the magnetic field extends from the photosphere at one end of the loop
to the photosphere at the other end. Let $s$ be the coordinate along the loop with $s = 0$
at the base of the photosphere at the positive polarity end of the loop, and $s = L$ at the
other end. From now on the field strength $B_0 (s)$, width $w_t (s)$ of a flux tube,
and width $w_m (s)$ of the merged field will be written as functions of $s$, not height in the
photosphere.
The points where the flux tubes merge are located at $s_1 = z_m$ and $s_2 = L- z_m$,
so the merged field extends from $s_1$ to $s_2$. This includes the coronal part of the loop,
as well as the chromospheres and TRs at the two ends. The merged field is approximated
as the potential field of a dipole located at some depth $z_0$ below the photosphere.
We use a cartesian coordinate system $(x^\prime,y^\prime,z)$ with the origin at the center
of the bipolar region and $z$ the height above the photosphere. The dipole is located at
$x^\prime = y^\prime = 0$, $z = -z_0$, and is pointed in the $-x^\prime$ direction. We also
use a spherical coordinate system $(r,\theta,\phi)$ centered at the dipole, where $r$ is the
radial distance from the dipole, $\theta$ is the angle relative to the $+x^\prime$ axis, and
$\phi$ is the azimuth angle ($\phi = 0$ corresponds to $y^\prime = 0$ in the cartesian frame).
Then the magnetic field is given by
\begin{equation}
B_r = - \mu \frac{2 \cos \theta} {r^3} , ~~~~
B_\theta = - \mu \frac{\sin \theta} {r^3} ,
\end{equation}
where $\theta$ is the angle relative to the $+x^\prime$ direction, and $\mu$ is the dipole
strength ($\mu > 0$). The field lines lie in planes $\phi$ = constant, and their shapes
are given by $r(\theta) = r_0 \sin^2 \theta$, where $r_0$ is constant along a field line.
For field lines in the plane $y^\prime = 0$ we obtain the following cartesian coordinates
\begin{eqnarray}
x^\prime(\theta) & = & r_0 \sin^2 \theta \cos \theta ,  \label{eq:xline} \\
z(\theta) & = & r_0 \sin^3 \theta - z_0 . \label{eq:zline}
\end{eqnarray}
We now consider the particular field line that intersects the photosphere at right angles,
which implies $dx^\prime /d\theta = 0$ at the two intersection points. Let $\theta (s)$
be the angle as function of position along this particular field line. Then equation
(\ref{eq:xline}) yields $\theta (0) = \pi - \theta_0$ and $\theta (L) = \theta_0$, where
$\theta_0 \equiv \arccos (1 / \sqrt{3})$, and equation (\ref{eq:zline}) yield
$r_0 = (3/2)^{3/2} z_0$. The intersection points are located at $x^\prime = \mp x_0$,
where $x_0 = z_0 / \sqrt{2}$. We assume $x_0 = 30$ Mm, which implies
$z_0 = 42.43$ Mm and $r_0 = 77.95$ Mm. The length of the loop between intersection
points is obtained by numerical integration: $L = 103.3$ Mm.
In this paper we consider a thin coronal loop surrounding the selected field line.
The magnetic field strength along this loop is given by
\begin{equation}
B (s) = | {\bf B} | = \frac{\mu} {r_0^3} \frac{ \sqrt{3 \cos^2 \theta (s) + 1}} {\sin^6 \theta (s)} ,
~~~~ \hbox{where} ~~ \theta_0 \le \theta (s) < \pi - \theta_0 .  \label{eq:B0}
\end{equation}
The dipole strength $\mu$ is chosen such that the average field strength along the
loop is 60 G. Then the field strength at the base of the photosphere is $B (0) = B (L)
= 181.6$ G. The field strength at the loop top is about  38 G.
From the photosphere to the loop top the cross-section of the loop expands by
a factor $\Gamma = (3/2)^3 \sqrt{2} = 4.77$.

\subsection{Wave Dynamics}
\label{wave}

We simulate the dynamics of transverse waves inside the flux tubes (photosphere)
and in the merged field (chromosphere and corona). The two regions are coupled such
that waves can propagate from the flux tubes into the merged field and back. In the
following we describe (a) the waves inside the flux tubes, (b) the waves in the merged field,
and (c) the reduced MHD equations describing these waves. The coupling between the
flux tubes and the merged field is described in the next subsection.

In this section we only consider the motions of plasma inside the photospheric flux tubes,
not the transverse motions of the tubes themselves. The side boundaries of the flux
tubes are assumed to have fixed positions. Therefore, we only simulate the Alfv\'{e}n
waves inside the flux tubes, not the kink waves that distort the shapes of the tubes.
The flux tubes have square cross-sections. For each flux tube we introduce coordinates
$(x,y)$ in the planes perpendicular to the tube axis; these coordinates are in the range
$-w_t /2 \le x \le +w_t /2$ and $-w_t /2 \le y \le +w_t /2$, where $w_t (s)$ increases with
height above the photosphere. The velocity inside the flux tubes is assumed to be
perpendicular to the flux tube axis, ${\bf v} (x,y,s,t) = v_x \hat{\bf x} + v_y \hat{\bf y} $,
where $\hat{\bf x}$ and $\hat{\bf y}$ are unit vectors in the horizontal direction. 
The velocity is assumed to be nearly incompressible, $\nabla_\perp \cdot {\bf v} = 0$,
and vanishes at the side boundaries of the tube: $v_x = 0$ at $x = \pm w_t (s) /2$, and
$v_y = 0$ at $y = \pm w_t (s)/2$. At the base of the photosphere the velocities
${\bf v} (x,y,0,t)$ or ${\bf v} (x,y,L,t)$ are imposed as boundary conditions (separately
for each flux tube). These so-called ``footpoint" motions vary randomly with time (see
section \ref{footpoint}), and are statistically independent for different flux tubes.
The footpoint motions produce Alfv\'{e}n waves that travel upward along the tubes.
When the waves reach the merging height they propagate into the merged field above.

At the merging height $4 \times 4$ flux tubes with square cross-sections come together to form
the merged field (see Figure~\ref{fig1}). Therefore, the merged field also has a square cross-section,
and the width $w_m (s)$ of the cross-section varies with position along the loop. At the merging
heights $w_m (s_1) = w_m (s_2) = $ 4,660 km. We introduce coordinates $(x,y)$ perpendicular
to the loop axis; these coordinates are in the range $0 \le x \le w_m (s)$ and $0 \le y \le w_m (s)$.
The velocity is  given by ${\bf v} (x,y,s,t) = v_x \hat{\bf x} + v_y \hat{\bf y}$, where $\hat{\bf x}$ and
$\hat{\bf y}$ are now perpendicular to the loop axis, which is curved (see previous subsection).
The velocity is again assumed to be nearly incompressible, $\nabla_\perp \cdot {\bf v} = 0$.
For the merged field we assume periodic boundary conditions, so ${\bf v} (x+w_m,y,s,t) =
{\bf v} (x,y+w_m,s,t) = {\bf v} (x,y,s,t)$, and similar for the magnetic fluctuations. 
The waves injected at the merging heights can travel upward along the loop and dissipate
their energy in the chromosphere and corona via AWT. The waves can also travel back down
into the flux tubes, and generate turbulence there. Therefore, the merged-field region contains
a complex wave field generated by waves from multiple flux tubes at each end of the coronal
loop.

The boundaries between the flux tubes at the merging height can be traced upward into the
merged field, forming separatrix surfaces. Away from the merging heights these boundaries
are no longer fixed, and can move with the transverse displacements of waves. Therefore,
within the merged field wave energy can be exchanged between neighboring flux tubes.
The velocity field may not be continuous at these boundaries because the waves on either
side originate in different photospheric flux tubes and are statistically independent.
Therefore, thin current sheets may develop at the boundaries between the flux tubes in
the merged field. The dissipation of these boundary currents may provide an extra source of
heat (in addition to the heat provided by AWT in a single tube).

The transverse waves are simulated using the reduced MHD approximation
\citep[e.g.,][]{Strauss1976, Strauss1997}. The magnetic field strength $B_0 (s)$ and density
$\rho (s)$ are assumed to be constant over the cross-section both for the flux tubes and for
the merged field. The magnetic and velocity fluctuations are described as a superposition
of Alfv\'{e}n waves propagating parallel and anti-parallel to the background field.
The waves can be written in terms of Elsasser variables,
\begin{equation}
{\bf z}_{\pm} (x,y,s,t) = {\bf v} \mp \frac{ {\bf B}_1 } {\sqrt{4 \pi \rho} } ,
\end{equation}
where ${\bf v} (x,y,s,t)$ is the plasma velocity, ${\bf B}_1 (x,y,s,t)$ is the magnetic
fluctuation, and $\rho (s)$ is the mean plasma density \citep[][]{Elsasser1950}.
The two different wave types interact nonlinearly: the ${\bf z}_{+}$ waves are distorted
by the counter-propagating ${\bf z}_{-}$ waves, and vice versa, which leads to a turbulent
cascade of wave energy to smaller spatial scales \citep[][]{Iroshnikov1963,
Kraichnan1965, Shebalin1983}. The Alfv\'{e}n waves are described in terms of stream
functions $f_{\pm} (x,y,s,t)$ for the Elsasser variables:
\begin{equation}
{\bf z}_{\pm} (x,y,s,t) = \nabla_\perp f_{\pm} \times \hat{\bf s} ,
\end{equation}
where $\hat{\bf s} (x,y,s)$ is the unit vector along the background field. Then the velocity
stream function $f = (f_{+} + f_{-})/2$, and the magnetic flux function $h = (f_{-} - f_{+})/
(2 v_{\rm A})$, where $v_{\rm A} (s) \equiv B_0 / \sqrt{4 \pi \rho}$ is the Alfv\'{e}n speed.
The reduced MHD equations \citep[e.g.,][papers~I and IV]{Strauss1976, Schekochihin2009,
Perez2013} can be written in the following form:
\begin{equation}
\frac{\partial \omega_{\pm}} {\partial t} = \mp v_{\rm A} \frac{\partial \omega_{\pm} }
{\partial s} + \frac{1}{2} \frac{dv_{\rm A}} {ds} \left( \omega_{+} - \omega_{-} \right)
+ {\cal N}_{\pm} + \tilde{\nu}_{\pm} \nabla_\perp^2 \omega_{\pm} ,   \label{eq:rmhd1}
\end{equation}
where $\omega_{\pm} \equiv - \nabla_\perp^2 f_{\pm}$ are the vorticities of the waves,
${\cal N}_{\pm}$ are nonlinear terms, and $\tilde{\nu}_{\pm}$ are artificial viscosities.
The nonlinear terms are given by
\begin{equation}
{\cal N}_{\pm} = - \onehalf [ \omega_{+} , f_{-} ] - \onehalf [ \omega_{-} , f_{+} ] 
\pm \nabla_\perp^2 \left( \onehalf [ f_{+} , f_{-} ] \right) 
\end{equation}
where $[ \cdots , \cdots ]$ is the bracket operator:
\begin{equation}
[a,b] \equiv \frac{\partial a} {\partial x} \frac{\partial b} {\partial y}
- \frac{\partial a} {\partial y} \frac{\partial b} {\partial x} ,
\label{eq:bracket1}
\end{equation}
where $a(x,y)$ and $b(x,y)$ are two arbitrary functions.
The four terms on the right-hand side of equation (\ref{eq:rmhd1}) describe wave propagation,
linear couplings resulting from gradients in Alfv\'{e}n speed, nonlinear coupling between
counter-propagating waves, and wave damping. The numerical methods for solving these
equations are described in Appendix~A. We use Fourier analysis to describe the dependence
of the waves on the $x$ and $y$ coordinates, and finite-differences in the $s$ direction along
the loop. To properly resolve the structure of the TR, we use a high-resolution grid with variable
grid spacing (the smallest cells have $\Delta s = 0.3$ km). Also, to follow the waves as they
propagate through the TR, we use very small time steps locally within the TR ($\Delta t <$
0.001~ s). Therefore, we no longer treat the TR as a discontinuity, as we did in our earlier
work (papers~I, II and III). This allows us to more accurately evaluate the heating rates within
the TR.

\subsection{Wave Coupling between Flux Tubes and Merged Field}
\label{coupling}

Consider a wave $f_{+}$ traveling upward in one of the flux tubes just below the
merging height at $s = s_1$. When this wave reaches the merging height it can readily
enter into the merged-field region, because the wave travels from the
relatively narrow flux tube into the much wider merged field. Hence, there is no
reflection of the wave as it reaches the merging level. However, the same is not true 
for waves traveling downward towards $s_1$. When a downward propagating wave
$f_{-}$ reaches the merging height, its transmission or reflection depends on the
perpendicular wavenumber $k_\perp$ of the wave. If the perpendicular length scale
$\pi / k_\perp$ of the wave is small compared to the width $w_t (s_1)$ of the flux tubes,
the wave can be readily transmitted, but if $\pi / k_\perp > w_t (s_1)$ the wave must be
partially reflected because the velocity field of the wave cannot satisfy the side
boundary conditions of the flux tubes.

The transmission and reflection of downward propagating waves at the merging height
are determined as follows. We first compute
the stream functions $f_{\pm}^\prime (x,y,s_1,t)$ of the waves at the bottom of
the merged field according to equation (\ref{eq:omegm}) (the grid refinement level
$N = 0$ at this height). Let $(x_b,y_b)$ be the coordinates of the edges of the flux
tubes, as indicated by the large squares in Figure 1. Also, let $f_b^\prime \equiv
f_{-}^\prime (x_b,y_b,s_1,t)$ be the values of the stream function for the $f_{-}$ wave
at these edges. We then compute the harmonic function $f^* (x,y,t)$ that satisfies
$\nabla_\perp^2 f^* = 0$ inside the squares and $f^* (x_b,y_b,t) = f_b^\prime$ at
the edges of the squares. This harmonic function does not satisfy the side boundary
conditions in the flux tubes, which require $f (x_b,y_b,t) = 0$. Therefore, we assume
that $f^* (x,y,t)$ is the part of the downward wave that is reflected back up into the
merged field (Longcope \& van Ballegooijen 2002). The remainder of the wave
function is given by
\begin{equation}
\delta f(x,y,t) \equiv f_{-}^\prime (x,y,s_1,t) - f^* (x,y,t) ,  \label{eq:df}
\end{equation}
which satisfies $\delta f (x_b, y_b,t) = 0$ and can be transmitted into the flux
tubes. The transmission is implemented by setting $f_{-} (x,y,s_1,t) = \delta f (x,y,t)$
at the top of the flux tubes just below the merging height. The reflection is
implemented by subtracting $f^* (x,y,t)$ from both wave types at the bottom of the
merged field:
\begin{equation}
f_{\pm} (x,y,s_1,t) = f_{\pm}^\prime (x,y,s_1,t) - \onehalf f^* (x,y,t) .
\end{equation}
The corrected stream function at the bottom of the merged field satisfies
\begin{equation}
f(x_b,y_b,s_1,t) = [f_{+} (x_b,y_b,s_1,t)+ f_{-}(x_b,y_b,s_1,t) ]/2 = 0 ,
\end{equation}
so the corrected velocity is parallel to the edges of the flux tubes, and is continuous
across the merging height, as required. A similar reflection occurs for the $f_{+}$ waves
at the other merging height, $s = s_2$.

\subsection{Photospheric Footpoint Motions}
\label{footpoint}

In the present model the Alfv\'{e}n waves are launched by imposing random footpoint motions
at the base of the photospheric flux tubes ($z = 0$) at both ends of the coronal loop.
Therefore, the velocity fields ${\bf v} (x,y,0,t)$ or ${\bf v} (x,y,L,t)$ are imposed as a boundary
conditions within the flux tubes. For each flux tube the velocity stream function at the base is
a sum of three modes:
\begin{equation}
f(x,y,t) = f_1 (t) F_{1,1} (\tilde{x},\tilde{y}) + f_2 (t) F_{2,1} (\tilde{x},\tilde{y})
+ f_3 (t) F_{1,2} (\tilde{x},\tilde{y}) ,
\end{equation}
where $\tilde{x} = x/w_0 +0.5$ and $\tilde{y} = y/w_0 + 0.5$ are dimensionless perpendicular
coordinates, and $F_{n_x,n_y} (\tilde{x},\tilde{y})$ is the eigenfunction defined in equation
(\ref{eq:Fk}). The three modes have $(n_x,n_y)$ = (1,1), (2,1) and (1,2), respectively, so the
first mode describes a rotational motion with a single cell, and the other modes each have
two counter-rotating cells. The mode amplitudes $f_k (t)$ are random functions of time. The
different driver modes within a flux tube are uncorrelated, and modes from different flux tubes
are uncorrelated. The random functions are obtained by Fourier filtering a random number
sequence, using the filter function $\exp [ - ( \tau_0 \nu )^2 ]$, where $\nu$ is the temporal
frequency (in Hz) and $\tau_0$ is a specified parameter. In this paper we use $\tau_0 = 180$~s,
which corresponds to a correlation time $\tau_c = \tau_0 / \sqrt{ 2 \pi } \approx 72$~s.
The filtered sequences are normalized such that each driver mode has an equal contribution
to the root-mean-square of the velocity. We assume $v_{\rm rms} = 1.5$ $\rm km ~ s^{-1}$,
similar to the value used in our earlier work (papers I, II and III).

\subsection{Construction of the Background Atmosphere}
\label{background}

In the Reduced MHD model the structure of the background atmosphere must be specified
before the waves can be simulated. In the lower atmosphere the temperature $T (s)$ and
mean molecular weight $\mu (s)$ are described using simple analytic models with a 
chromospheric temperature of 8000 K. The gas pressure $p(s)$ and density $\rho (s)$
in the lower atmosphere are computed from the hydrostatic equilibrium equation. In the
chromosphere the pressure scale height $H_p (s)$ is increased by $20\%$ to account for
wave pressure forces.

The temperature $T(s)$ and density $\rho (s)$ in the transition region and corona are
determined by solving the equations describing the energy balance of the coronal plasma
(for details, see Appendix~B). The energy equations (\ref{eq:loop}) and (\ref{eq:Fs}) include
the effects of wave heating, thermal conduction, enthalpy flux and radiative losses.
The methods for solving these equations are similar to those used in \citet[][] {Schrijver2005},
except that in the present case the coronal pressure
$p_{\rm cor}$ is treated as a free parameter of the model, and the heating rate $Q_{\rm A} (s)$
is a derived quantity. For the purpose of constructing the background atmosphere, we assume
the Alfv\'{e}n wave heating rate is of the form
\begin{equation}
Q_{\rm A} (s) = c_0 [ B_0 (s)]^n ,  \label{eq:QA}
\end{equation}
where $c_0$ is a constant and the power law exponent $n \ge 0$. The constant $c_0$ is
determined from $p_{\rm cor}$ as part of the iteration process (see Appendix~B). The radiative
loss function $\Lambda (T)$ is taken from CHIANTI version 8 \citep[][]{Dere1997, DelZanna2015a},
assuming coronal abundances \citep[][]{Schmelz2012}. The loop modeling yields $T (s)$ and
$\rho (s)$ as functions of position along the loop, as well as the heating rate $Q_{\rm A} (s)$
needed to maintain the assumed coronal pressure. The parameter $p_{\rm cor}$ also
determines the height $z_{\rm TR}$ of the base of the transition region.

\section{RESULTS FOR ALFV\'{E}N WAVE TURBULENCE MODEL}
\label{with_tubes}

In this section we describe results from 3D MHD simulations of Alfv\'{e}n waves for the
multi-flux-tube model presented in section \ref{multi}. The background atmosphere for this
model is constructed as follows. Using an estimated value for the coronal pressure
$p_{\rm cor}$, we set up the background atmosphere as described in section~\ref{background}
and then simulate the wave dynamics as described in section~\ref{wave}. The waves are
simulated for a period $t_{\rm max} = 3000$ s, which is sufficient to reach a statistically
stationary state where turbulence is present everywhere along the loop. From these simulation
results we derive the wave heating rate $Q_{\rm tot} (s)$, which is an average over the loop
cross-section and over time. In the corona $Q_{\rm tot} (s)$ is not necessarily equal to the rate
$Q_{\rm A} (s)$ assumed in the setup of the background atmosphere. We then adjust
$p_{\rm cor}$ and the exponent $n$ in equation (\ref{eq:QA}) until an approximate agreement
between the two heating rates is obtained. The final condition $Q_{\rm tot} \approx Q_{\rm A}$
indicates that the loop is in thermal equilibrium such that the simulated wave heating is
balanced by the radiative and conductive losses of the coronal plasma. We find that the
thermal equilibrium condition is satisfied when $n \approx 1$ and $p_{\rm cor} \approx 1.8$
$\rm dyne ~ cm^{-2}$. Note that this value of coronal pressure applies
only for the particular set of model parameters described in section \ref{multi}.
For example, we assumed that the photospheric magnetic flux density $\overline{B_z} =
181.6$ G, the photospheric footpoint velocity $v_{\rm rms} = 1.5$ $\rm km ~ s^{-1}$, and
the coronal loop length $L_{\rm c} = 98.4$ Mm. In the following we show results for
(1) the properties of the background atmosphere,
(2) the time-averaged wave properties, (3) the magnetic- and velocity structure of the waves,
(4) the spatial and temporal variations of the heating, and (5) the effect of the waves on
spectral line profiles.

\subsection{Background Atmosphere}

\begin{figure}
\centering
\includegraphics[width=6in]{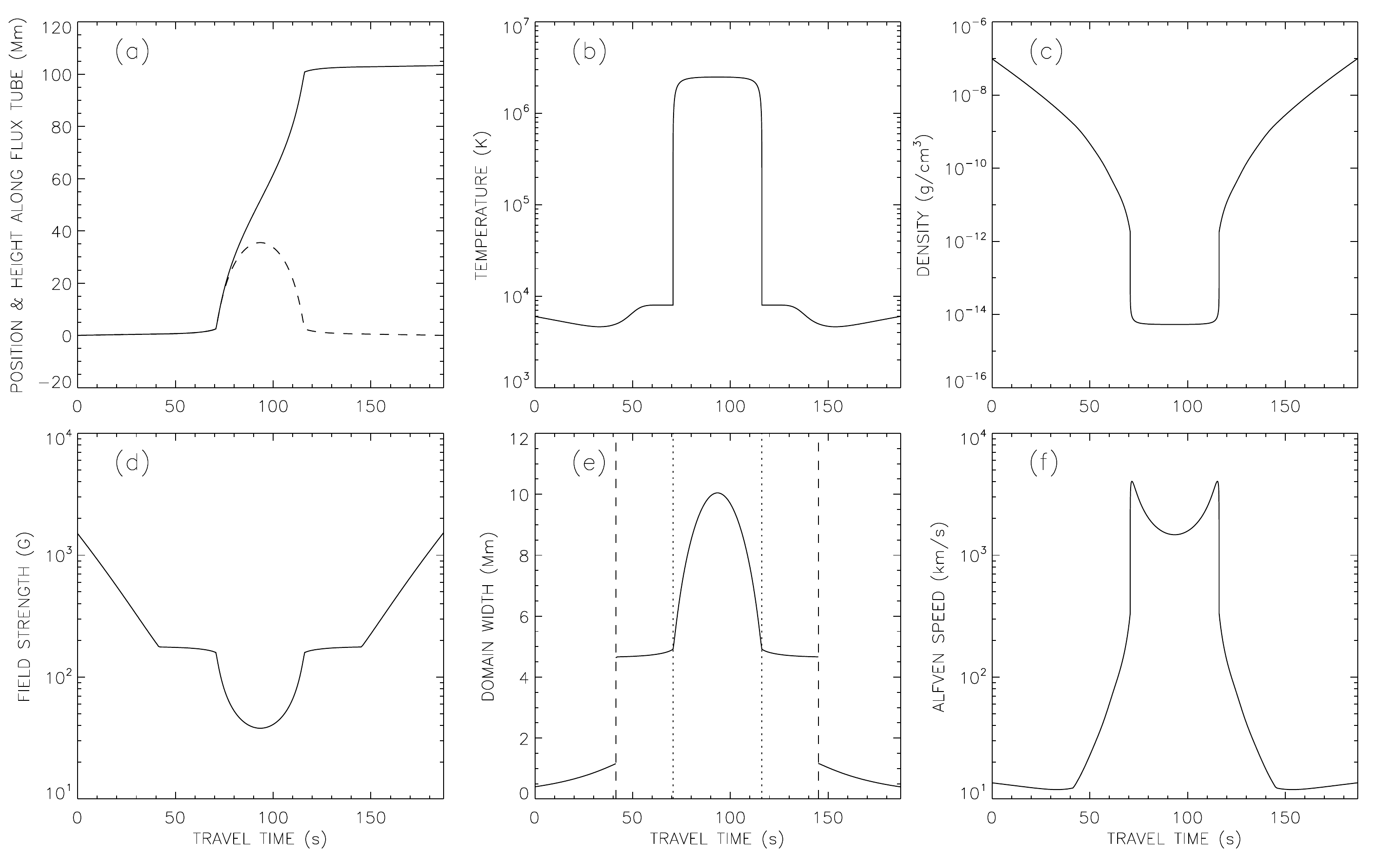}
\caption{Parameters of the background atmosphere in a coronal loop. The parameters are
plotted as function of the position along the loop, which is expressed in terms of the Alfv\'{e}n travel
time $t_0 (s)$ from the positive-polarity footpoint. (a) Position $s$ along the loop (full curve)
and height $z$ above the photosphere (dashed curve). (b) Temperature $T(s)$. (c) Mass density
$\rho (s)$. (d) Magnetic field strength $B_0 (s)$. (e) Width $w_t (s)$ of the individual flux tubes
(for $t_0 < 41$ s, and $t_0 > 145$ s), and width $w_m (s)$ of the merged field (for $41 < t_0 < 145$ s).
The dashed and dotted lines indicate the merging heights and TRs, respectively.
(f) Alfv\'{e}n speed $v_{\rm A} (s)$.}
\label{fig2}
\end{figure}

The properties of the background atmosphere
are shown in Figure~\ref{fig2}. Here various quantities are plotted as function of position
along the loop (merged field and flux tubes). Since the Alfv\'{e}n speed varies strongly with
position, it is convenient to use the wave travel time $t_0 (s) \equiv \int_0^s ds^\prime /
v_A (s^\prime)$ as the independent variable. In terms of this coordinate the two merging heights
are located at 41.4~s and 145.0~s, the two TRs are located at about 70~s and 117~s, the loop top
is located at 93.2~s, and the total wave travel time along the loop is 187.1~s. Figure~\ref{fig2}(a)
shows the position $s$ (full curve) and height $z$ (dashed curve) as functions of $t_0$.
Note that the total loop length is about 103.3 Mm, and the height of the loop top is
35.5~Mm. In Figure~\ref{fig2}(b) the temperature $T$ is plotted as function of $t_0$, which
causes the lower atmosphere ($T < 10^4$~K) to be greatly expanded compared to the corona
($T > 10^6$~K). The temperature plateaus with $T = 8000$ K in the chromosphere are located
at $50 < t_0 < 70$~s and $117 < t_0 < 137$~s. In the corona the peak temperature
$T_{\rm max} = 2.496$ MK, which is less than the values of  3 -- 4 MK found at the peaks
of the observed DEM distributions \citep[e.g.,][]{Winebarger2011, Warren2011, Warren2012}.
Figure~\ref{fig2}(c) shows the plasma density $\rho$, which varies over 7 orders of magnitude.

The magnetic field strength $B_0$ along the loop is shown in Figure~\ref{fig2}(d). Note that
the horizontal scale of the plot is strongly distorted by the fact that we plot $B_0$ as function of
wave travel time $t_0$, not distance $s$. Also note that $B_0$ is continuous at the merging
height, and the minimum field strength of 38 G occurs at the loop top. Figure~\ref{fig2}(e) shows the
widths $w_t (s)$ of the individual flux tubes, and the width $w_m (s)$ of the merged field. Here
we also plot the locations of merging heights (dashed vertical lines) and TRs (dotted lines).
Finally, Figure~\ref{fig2}(f) shows the Alfv\'{e}n speed $v_{\rm A}$. Note that $v_{\rm A}$
increases from about 15 $\rm km ~ s^{-1}$ in the photosphere to about 4000 $\rm km ~ s^{-1}$
in the low corona, then drops to 1500 $\rm km ~ s^{-1}$ at the loop top. The rise of $v_{\rm A}$
in the chromosphere and TR causes strong wave reflection. The resulting counter-propagating
waves interact nonlinearly and produce AWT in the lower atmosphere (paper~I).

\subsection{Time-Averaged Wave Properties}

\begin{figure}
\centering
\includegraphics[width=6in]{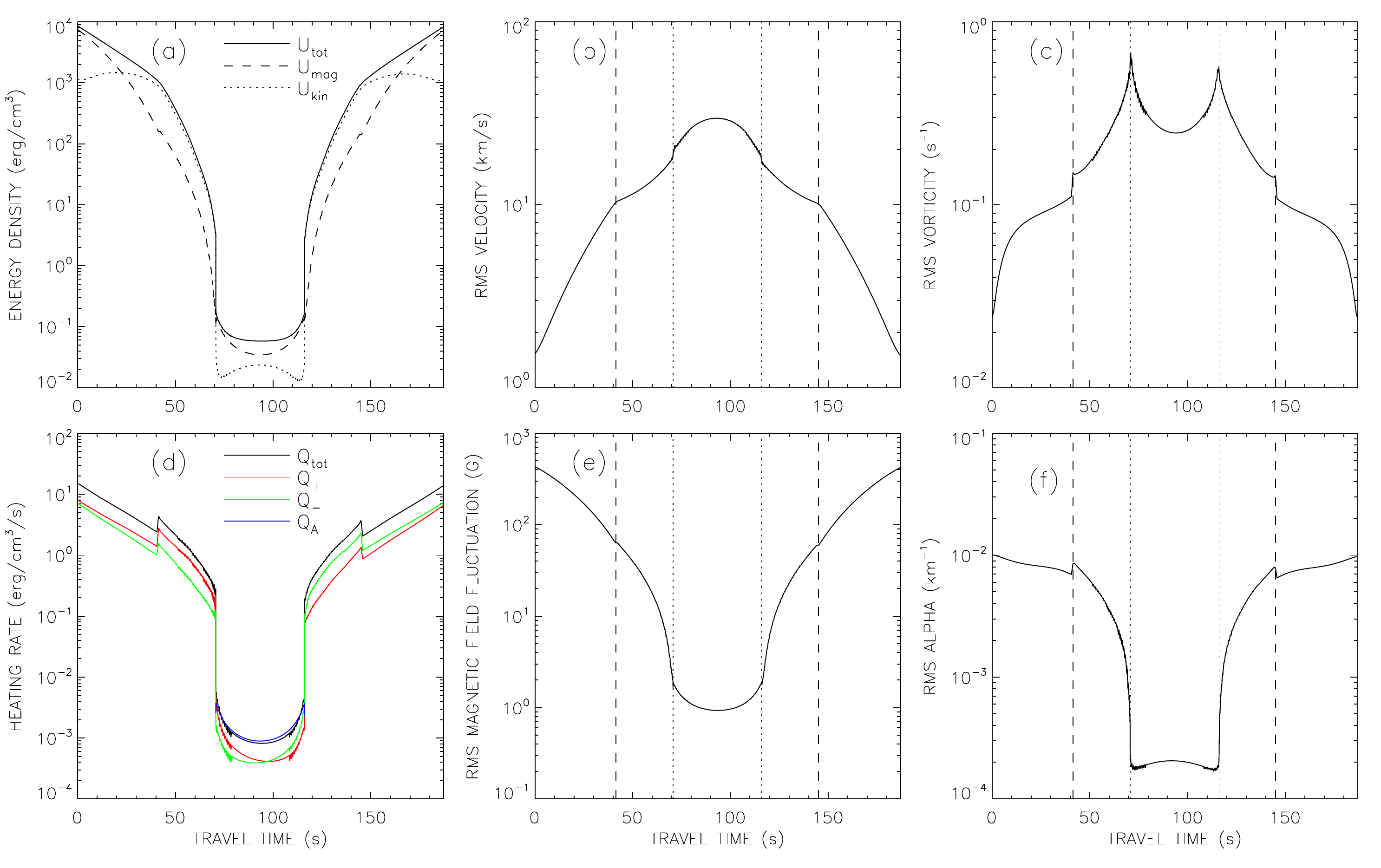}
\caption{Model for Alfv\'{e}n wave turbulence in a coronal loop. Various wave-related parameters
are plotted as function of the position along the loop, which is expressed in terms of the
Alfv\'{e}n travel time $t_0 (s)$ from the positive-polarity footpoint. (a) Kinetic and magnetic energy
densities, and their sum $U (s)$. (b) Velocity amplitude $v_{\rm rms} (s)$ of the waves.
(c) Vorticity amplitude $\omega_{\rm rms} (s)$. (d) Wave energy dissipation rates per unit volume:
total wave dissipation rate $Q_{\rm tot} (s)$ (black curve), rate $Q_{+}$ for waves traveling from
left to right (red curve), rate $Q_{-}$ for waves traveling from right to left (green curve), and plasma
heating rate $Q_{\rm A}$ assumed in the setup of the background atmosphere (blue curve).
(e) Amplitude of magnetic field fluctuations, $B_{1, \rm rms}$. (f) Magnetic twist parameter,
$\alpha_{\rm rms}$. In panels (b), (c), (e) and (f), the dashed and dotted lines indicate the
merging heights and TRs, respectively.}
\label{fig3}
\end{figure}

Figure~\ref{fig3} shows various wave-related quantities as function of position along the loop,
averaged over time $t$ in the simulation ($200 < t < 3000$ s). In the photosphere these
quantities are also averaged over the cross-sections of all flux tubes combined, and in the
chromosphere and corona they are averaged over the cross-section of the merged field. 
Figure~\ref{fig3}(a) shows the total energy density of the waves, and the contributions from
magnetic- and kinetic energy. Note that in the chromosphere the kinetic energy dominates,
while in low corona ($t_0 = 75$ s and $t_0 = 110$ s) the magnetic energy is more important.
Figure~\ref{fig3}(b) shows the velocity amplitudes $v_{\rm rms}$ of the waves; this quantity
peaks at the loop top, where $v_{\rm rms} \approx 30$ $\rm km ~ s^{-1}$. Figure~\ref{fig3}(c)
shows the vorticity, which peaks in the low corona. The maximum vorticity likely depends
on the resolution of the numerical model.

Figure~\ref{fig3}(d) shows the wave heating rate $Q_{\rm tot} (s)$ as predicted from the
numerical simulations (black curve), as well as the contributions to $Q_{\rm tot}$ from waves
propagating in the positive and negative $s$ directions (red and green curves, respectively).
Note that the heating rates in the corona are about $10^{-3}$ $\rm erg ~ cm^{-3} ~ s^{-1}$,
much smaller than those in the photosphere and chromosphere.  We also plot the heating
rate $Q_{\rm A} (s)$ used in the setup of the background atmosphere (blue curve). Note
that $Q_{\rm tot} \approx Q_{\rm A}$, which is due to our choice for the coronal pressure,
$p_{\rm cor} = 1.8 $ $\rm dyne ~ cm^{-2}$. Looking at this figure more closely, we see that
there are small jumps in $Q_{\rm tot}$ at the merging heights ($t_0 = 41.4$ s and
$t_0 = 145.0$ s). These jumps indicate that the average heating rate in the region just above
the merging height $z_m$ is larger than that in the flux tubes just below $z_m$. As we will
show later, this extra heating is due to electric currents at the interfaces between flux tubes in
the chromosphere, where the tubes first come together. Figure~\ref{fig3}(e) shows the amplitude
$B_{1,\rm rms}$ of the magnetic field fluctuations. The largest fluctuations occur in the lower
atmosphere. At the loop top $B_{1,\rm rms} \approx 1$~G, which is small compared to the local
field strength, $B_0 = 38$~G. Figure~\ref{fig3}(f) shows the rms value of the twist parameter,
$\alpha \equiv ( \nabla \times {\bf B} )_\parallel / | {\bf B} |$.
Note that in the coronal part of the loop the twist parameter is more or less constant,
$\alpha_{\rm rms} \approx 2 \times 10^{-4}$ $\rm km^{-1}$, indicating that the coronal field
is close to a force-free state. However, in the lower atmosphere much larger values of
$\alpha_{\rm rms}$ are found, so the global magnetic field is far from a force-free state.
This is a consequence of the fact that the photospheric footpoint motions produce wave-like
disturbances with most of the inertia of the waves located in the lower atmosphere.

\subsection{Magnetic- and Velocity Structure of the Waves}

\begin{figure}
\centering
\includegraphics[width=7in]{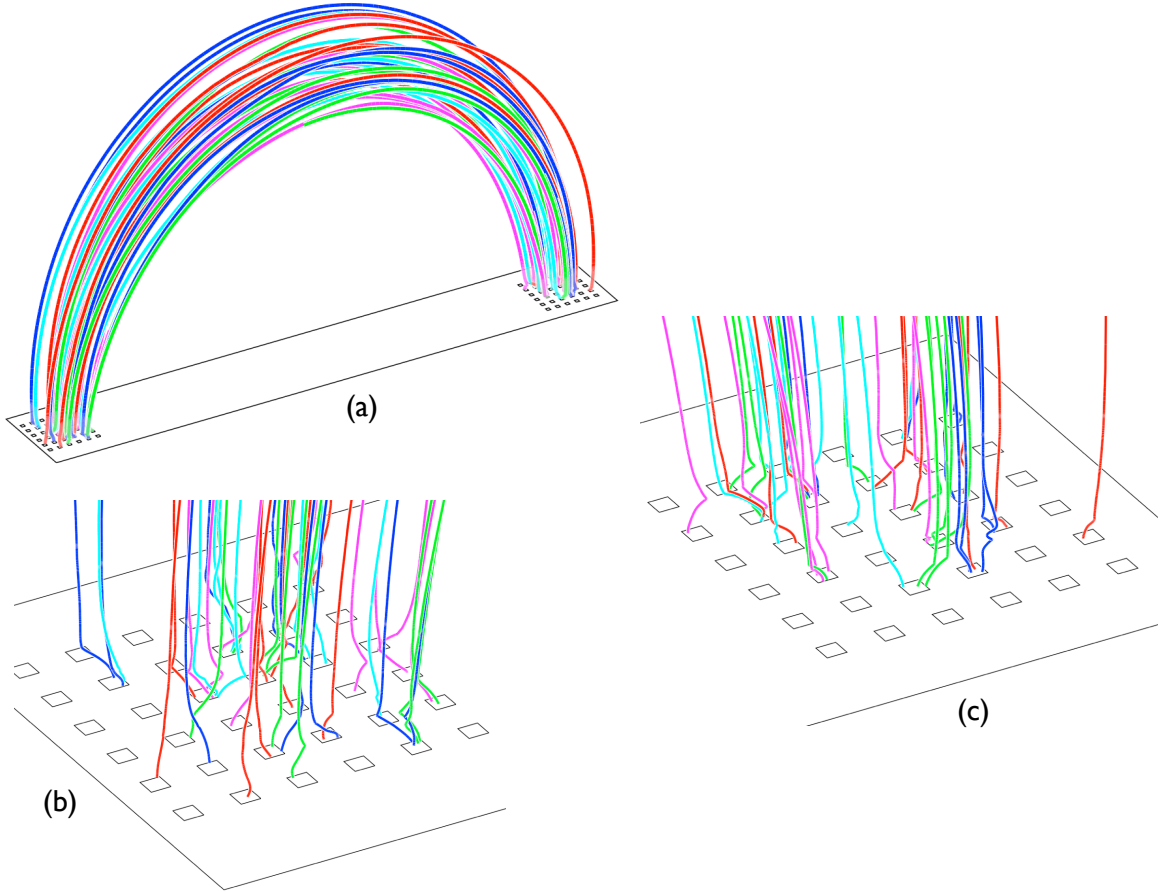}
\caption{Magnetic field lines in a coronal loop heated by AWT. The colors of the field lines are
randomly selected. (a) Large-scale view showing the coronal part of the loop. Note that the
deviations from the (dipole) potential field are very small. (b) Close-up of the left footpoint,
where the magnetic field breaks up into an array of discrete flux tubes. The squares indicate
the intersections of the flux tubes with the plane $z = 0$. (c) Similar close-up of the right footpoint.}
\label{fig4}
\end{figure}

The magnetic structure of the loop is illustrated in Figure~\ref{fig4}, where we plot the shapes
of the magnetic field lines at the end of the simulation ($t = 3000$~s). The field lines are started
from 48 randomly selected points in the cross-section at the loop top, and are traced downward
along the legs of the loop and into the flux tubes at the two ends. The field-line tracing takes into
account the overall curvature of the loop, the variation of the background field ${\bf B}_0 ({\bf r})$,
and the variation of the perturbations ${\bf B}_1 ({\bf r})$. We use periodic boundary
conditions at the side walls, so the field lines are not confined to the computational domain but
may pass into neighboring regions. Figure~\ref{fig4}(a) shows the field lines on the scale of the loop.
Note that the field lines have small tilt angles relative to the loop axis, consistent with the fact that
in our model $B_{1,\rm rms} \ll B_0$ in the coronal part of the loop. The tilt angles are only a few
degrees, much smaller than the angles predicted for nanoflare models \citep[e.g.,][] {Parker1983,
Pontin2015}. However, such small angles appear to be consistent with the observed fine
structures of coronal loops, which show only small deviations from the direction of the mean
magnetic field \citep[][]{Schrijver1999, Antolin2012, Brooks2013, Scullion2014}.
Also, the fact that these angles are small is consistent with our use of the reduced MHD
approximation. Figures~\ref{fig4}(b) and \ref{fig4}(c) show close-ups of the two ends of the loop,
where the field lines enter into the flux tubes (as sketched in Figure~\ref{fig1}). The small squares
indicate the intersections of the flux tubes with the base of the photosphere ($z = 0$). Note that
there are small kinks in the field lines at the merging height. These kinks are due to the discontinuity
of the horizontal components of the background magnetic field at the merging height.

\begin{figure}
\centering
\includegraphics[width=6.5in]{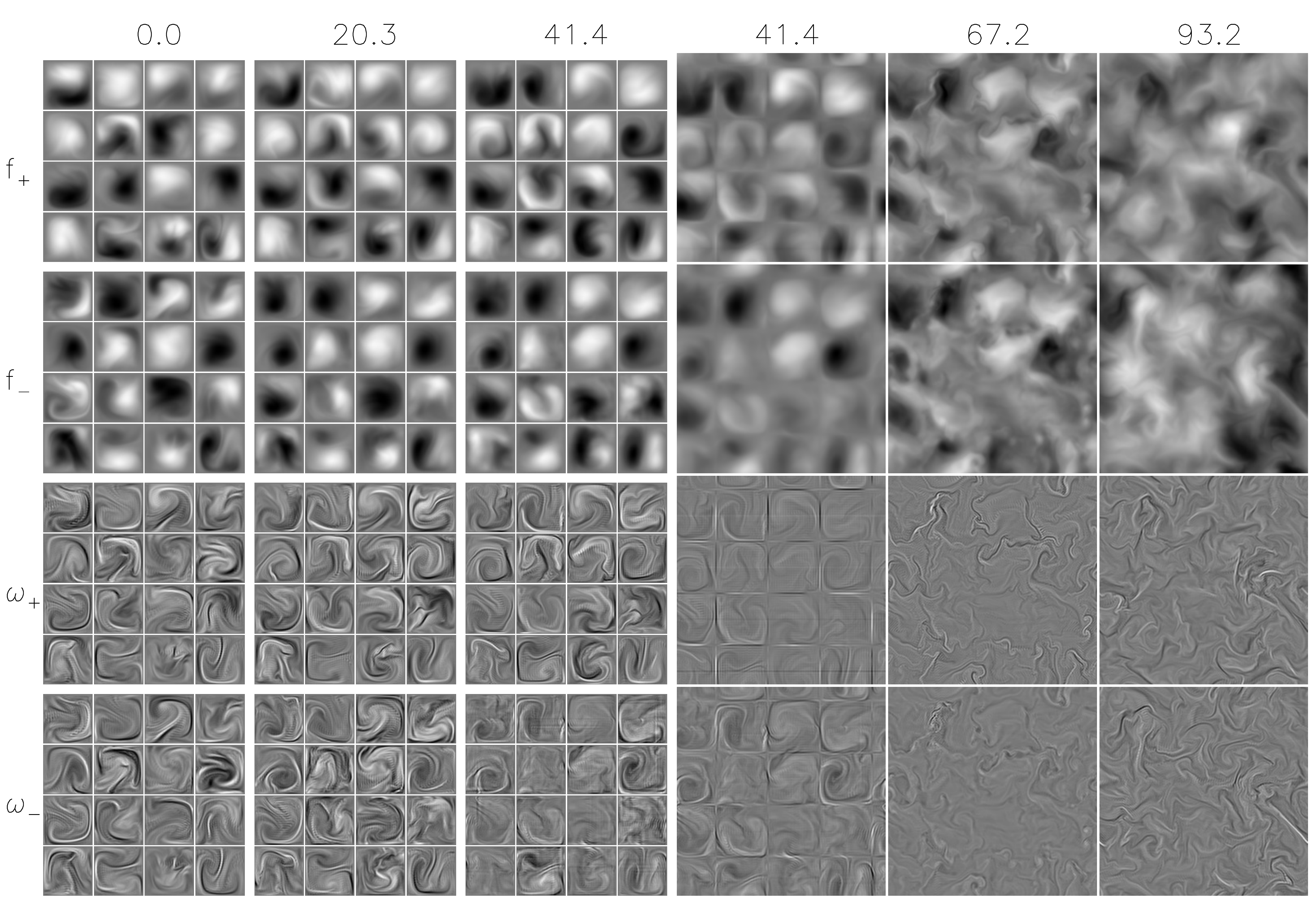}
\caption{Spatial distribution of the Elsasser variables in a 3D MHD model for AWT in
a coronal loop: velocity stream functions $f_{\pm} (x,y)$ (top rows) and vorticities
$\omega_{\pm} (x,y)$ (bottom rows). The different columns correspond to cross-sections at
different positions along the loop, and are labeled with the Alfv\'{e}n travel time $t_0 (s)$.
The left three columns show cross-sections of the flux tubes in the photosphere at the positive
polarity end of the loop, and the remaining columns show cross-sections of the merged field.
The merging height is shown twice ($t_0 = 41.4$ s), and $t_0 = 93.2$ s corresponds to the loop
top. The quantities $f_{\pm}$ and $\omega_{\pm}$ are shown as normalized greyscale images
with white (black) indicating positive (negative) signals (see Figure~\ref{fig3} for information
about the normalization and the actual width of each image).}
\label{fig5}
\end{figure}

Figure~\ref{fig5} shows the structure of the waves in cross-sections of the loop at time
$t = 3000$~s. The diagram has four main rows: the first and second rows show the stream
functions of the Elsasser variables, $f_{\pm} (x,y)$, and the third and fourth rows show the
vorticities, $\omega_{\pm} (x,y)$, as functions of the perpendicular coordinates
$x$ and $y$. The figure also has six main columns, corresponding to different positions
along the loop as indicated by the wave travel time $t_0$ at the top of each column. The first
three columns show the waves in the flux tubes at the positive polarity end of the loop.
In these columns $f_{+}$ is the upward propagating wave, and $f_{-}$ is the downward
propagating wave. For these three columns the wave patterns are shown as an array of
$4 \times 4$ small images, each one corresponding to an individual flux tube (note that
$f_{\pm} = 0$ at the side boundaries of the flux tubes). The size of these small images is
normalized, and does not reflect the actual width $w_t (s)$ of the flux tubes. The last three
columns show the wave patterns in the merged field in the left leg of the loop. Again, the
size of these larger images does not reflect the actual width $w_m (s)$ of the merged field.

The merging height on the positive polarity side ($t_0 = 41.4$~s) is shown twice: in the third
column as a collection of flux tubes, and in the fourth column as a merged field. Comparing
the images for $f_{+} (x,y)$ in these two columns, we see that the waves in the flux tubes are
nearly identical to those in the merged field. The reason is that a grid pattern is imposed on the
upward propagating waves in the merged field, where $f_{+} \approx 0$ at the grid boundaries.
The stream function $f_{+}$ at these boundaries is not exactly zero because of wave reflection
at the merging height (see section \ref{coupling}). A similar grid pattern can be seen for the
downward propagating waves $f_{-}$ at the merging height, which is due to wave reflection
at larger heights in the chromosphere. This grid pattern is not visible in the upper chromosphere
($t_0 = 67.2$ s) and loop top ($t_0 = 93.2$ s), indicating that the boundaries between flux tubes
have little effect on the waves at larger heights. The third and fourth rows of Figure~\ref{fig5} show
the vorticities $\omega_{\pm} (x,y)$ of the Elsasser variables at the same positions along the loop.
The actual vorticity is given by $\omega = (\omega_{+} + \omega_{-})/2$, and the current density
is proportional to the twist parameter, $\alpha = (\omega_{-} - \omega_{+})/ (2 v_{\rm A})$. Note
that the vorticity is concentrated in thin layers where the magnetic field ${\bf B}_1$ and velocity
field ${\bf v}$ change rapidly with position. Such current sheets and shear layers are naturally
produced by the turbulent flow, and are further enhanced by discontinuities in the velocity at the
boundaries between flux tubes at the merging height.
\begin{figure}
\centering
\includegraphics[width=7in]{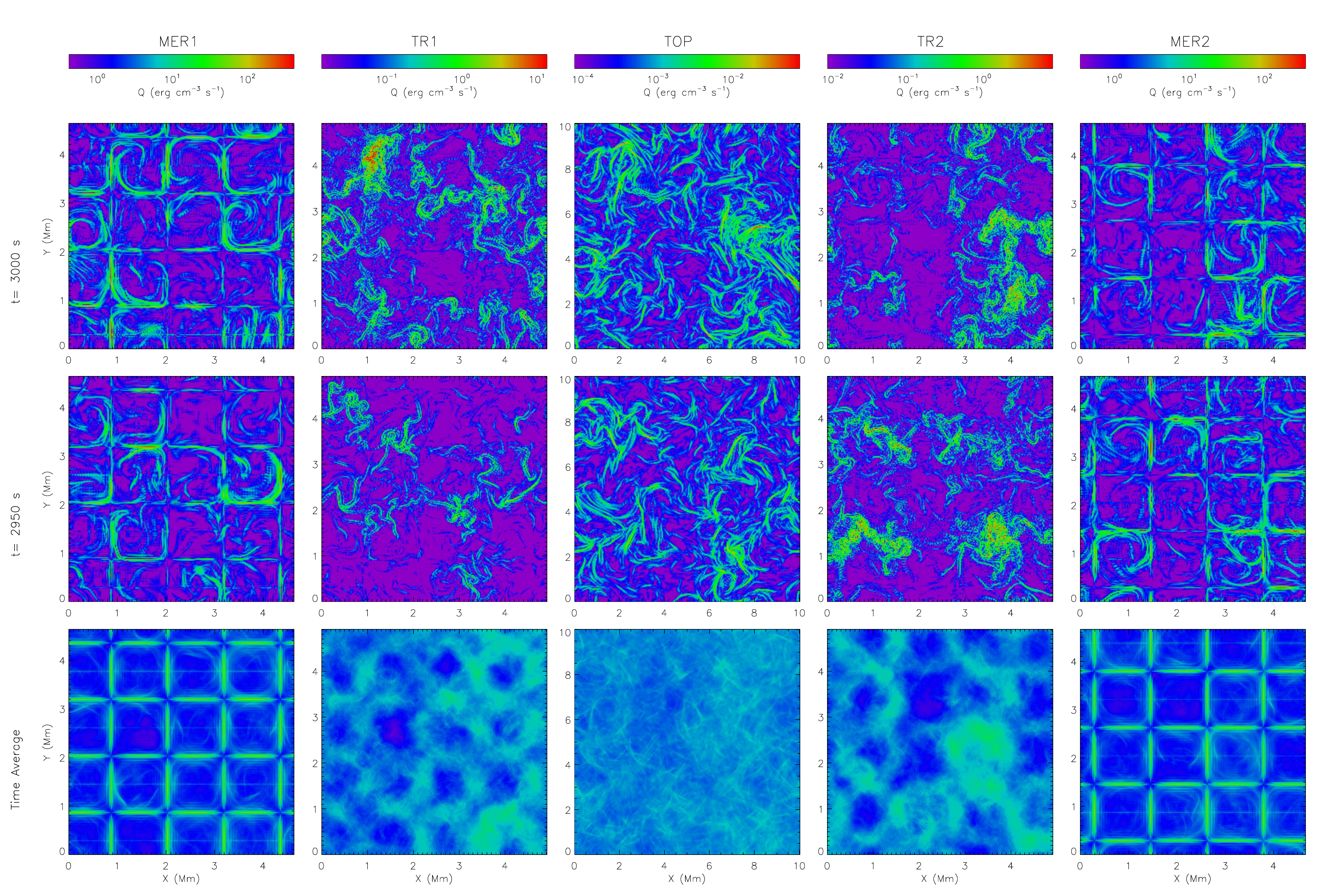}
\caption{Spatial distribution of the heating rates in a 3D MHD model for AWT in a coronal loop.
The top and middle rows show the heating at times $t = 3000$ s and $t = 2950$ s, respectively;
the bottom row shows the heating rate averaged over the time interval $200 < t < 3000$ s. The
different columns correspond to different positions along the loop, as indicated by the labels at
the top of each column: MER1 and MER2 are the merging heights, TR1 and TR2 are the
chromosphere-corona transition regions, and TOP is the loop top. Each image shows the rate
$Q(x,y)$ as a function of the perpendicular coordinates $x$ and $y$, and normalized as shown
by the color bar at the top of each column. Note that the color scales are logarithmic.}
\label{fig6}
\end{figure}

Note that the grid pattern in the merged field in the fourth column of Figure~\ref{fig5} is shifted
relative to the pattern of the flux tubes in the third column. The shift is to the lower left, and its
magnitude is one quarter of the width of a flux tube, or $w_t (z_m)/4$, in both the $x$ and $y$
directions. A similar shift occurs in the other leg of the loop (not shown), such that there is an
overall shift of $w_t (z_m)/2$ between the flux tubes at the two ends. Therefore, the flux tubes
at the two ends of the loop are not aligned with each other.

\subsection{Spatial and Temporal Variations of the Heating}
\label{variation}

We now consider the plasma heating produced by the waves. In the present model the wave
damping is described as ordinary diffusion, not hyper-diffusion as we did in our earlier work
(papers~I, II and III). The damping rates $\nu_{\pm,k}$ of the various wave modes are assumed
to be proportional to the square of the perpendicular wavenumber (see Appendix~A). This
allows us to compute the spatial distribution of the wave dissipation rate $Q(x,y,s,t)$ as function
of $x$ and $y$. The wave energy dissipation rate is assumed
to be equal to the heating rate of the plasma. Figure~\ref{fig6} shows the spatial distribution of
the heating at two different times near the end of the simulation (top and middle rows),
as well as its time average over the duration of the simulation (bottom row). The different
columns correspond to different positions along the loop, as indicated by the labels at the top
of each column. The heating rates are shown on a
logarithmic scale, but a different scale is used for each column (see color bars at the top).
Note that the instantaneous heating rates vary by more than 2 orders of magnitude over the
cross section of the loop for all heights. Most of the heating occurs in current sheets and shear
layers. At the merging heights the strongest heating occurs near the boundaries between
flux tubes, which have fixed positions. However, in the TR and corona the location of the
current sheets changes rapidly with time, so the spatial distribution of the heating also
changes rapidly (compare upper and middle rows, which have a time difference of only 50~s).
When the heating rates are averaged over time, the spatial variations across the loop become
much less pronounced (see bottom row). At the merging height the time-averaged rate shows
a strong grid pattern due to the underlying flux tubes, but the pattern is much weaker at the TRs,
and absent at the loop top. Therefore, the effect of the flux tubes does not extend into the corona,
and at the loop top the time-averaged heating rate is nearly uniform across the loop (see image
in middle column, bottom row).
\begin{figure}
\centering
\includegraphics[width=6in]{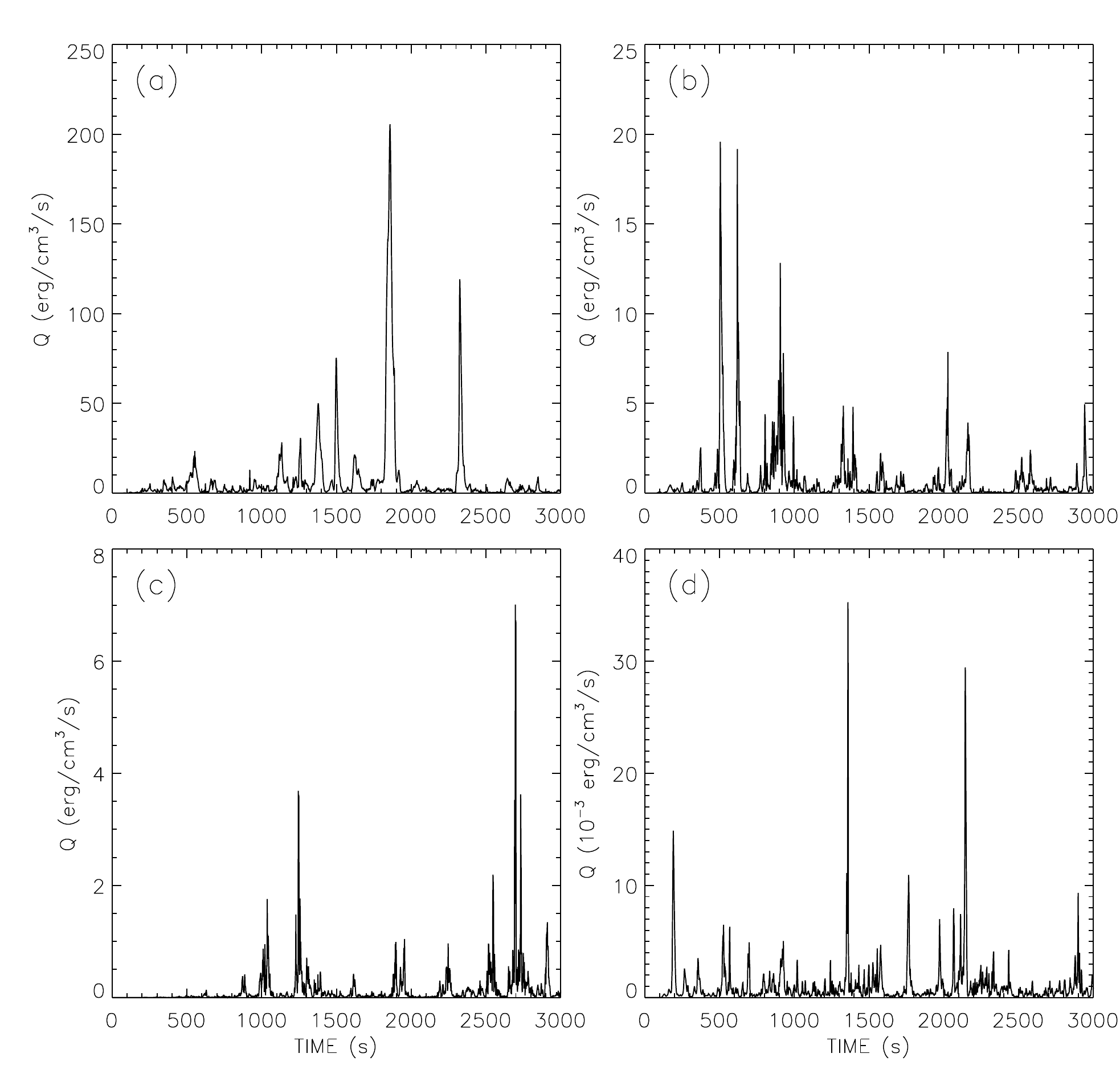}
\caption{Heating rate $Q(t)$ as function of time for random points moving with the flow at
different positions along the loop: (a) merging height; (b) mid-chromosphere, (c) transition region,
(d) loop top.}
\label{fig7}
\end{figure}

To understand how the plasma might respond to the heating, it is useful to determine how
the heating rate $Q(t)$ varies with time $t$ for a point moving with the flow. We randomly select
various points in different cross-sections of the merged field, and track these points for the
duration of the simulation. In the reduced MHD model the flows along the background field are
neglected, so each point remains in the cross-section where it was originally located,
but its perpendicular coordinates $x(t)$ and $y(t)$ vary with time. We measured the heating
rates $Q(t)$ at such co-moving points. The results are shown in Figure~\ref{fig7} for four points
at different positions along the loop. In all cases the heating is found to be very intermittent with
strong heating events superposed on a background of much weaker heating. The magnitude
of the heating events can be compared with the thermal energy density of the plasma,
$E_{\rm th} = (3/2) p$, where $p$ is the plasma pressure. Figure~\ref{fig7}(a) shows the heating
rate for a co-moving point at the merging height, $z_m = 518$~km. Note that
there are about 8 events with peak heating rates in excess of 20~$\rm erg ~ cm^{-3} ~ s^{-1}$.
Each event lasts about 20 -- 80 s and produces about 400 -- 4000 $\rm erg ~ cm^{-3}$,
comparable to the thermal energy density at that height, $E_{\rm th} \approx 823$
$\rm erg ~ cm^{-3}$. Therefore, such heating events are expected to produce significant
increases in the local temperature. Figure~\ref{fig7}(b)) shows the heating rate for
a point in the middle chromosphere ($z = 1696$ km), where the heating events are an order
of magnitude weaker. However, the thermal energy density is also much smaller,
$E_{\rm th} \approx 15.8$ $\rm erg ~ cm^{-3}$, so again we expect significant temperature
variations. The same is true for the base of the TR ($z = 2467$ km), where $E_{\rm th} \approx
2.7$ $\rm erg ~ cm^{-3}$ (see Figure~\ref{fig7}(c)). Finally, Figure~\ref{fig7}(d)
shows the heating rate for a co-moving point at the loop top. There are about 13 events with
peak heating rates in excess of 0.004~$\rm erg ~ cm^{-3} ~ s^{-1}$ and durations of about
20~s, so each event produces about 0.08~$\rm erg ~ cm^{-3}$. This is small compared to the
thermal energy density at the loop top, $E_{\rm th} \approx 2.7$ $\rm erg ~ cm^{-3} ~ s^{-1}$,
so the simulated heating events are not expected to produce large temperature
fluctuations at the loop top. This is consistent with earlier results in paper~II, where we
considered heating rates averaged over the loop cross-section.

The present AWT model predicts that the time-averaged heating rate $\overline{Q} (x,y,s)$
is enhanced at the boundaries between the flux tubes in the chromosphere, i.e., there is
extra heating at the separatrix surfaces in the chromosphere. In the region just above the
merging height, the time-averaged heating rate $Q_{\rm tot} (s)$ is enhanced by about
a factor 2 compared to its value just below the merging height. However, this extra heating
at the flux tube boundaries does not extend to large heights. At the height of the TR, the
boundary heating is relatively weak, and at the loop top the time-averaged rate 
$\overline{Q} (x,y,s)$ is more or less constant over the cross-section of the loop. The
reason for this rapid decline of the boundary heating with height is that the waves in the
chromosphere are quite turbulent, and produce large displacements of the separatrix
surfaces. Therefore, the extra energy associated with the discontinuities in the velocity at
the merging height is quickly dissipated.

\subsection{Effect of the Waves on Spectral Line Profiles}
\label{doppler}

Spectroscopic observations of coronal loops show that the observed emission lines are
significantly broadened in excess of their thermal widths \citep[e.g.,][]{Doschek2007,
Doschek2012, Young2007, Tripathi2009, Warren2011, Tripathi2011, Tian2011, Tian2012a,
Tian2012b}. Transverse MHD waves may contribute to this broadening, and at high
spatial resolution it may be possible to directly observe the Doppler shifts associated
with such waves. To compare the present model with observations, we simulate
the effect of the modeled waves on the Doppler shift and Doppler width of an observed
spectral line, assuming that the coronal loop is viewed from the side (in the $+y$ direction).
Then the line-of-sight (LOS) velocity is given by $v_y$, and the observed Doppler shift is
proportional to $< v_y >$, the mean value of $v_y$ along the LOS. Also, the non-thermal
component of the Doppler width is proportional to the velocity variance $\sigma_y$,
which is given by $\sigma_y^2 = < v_y^2 > - ( < v_y > )^2$.  We assume for simplicity
that the emissivity and thermal width of the spectral line are constant over the loop
cross-section. Then $< v_y >$ and $< v_y^2 >$ can be approximated as simple
averages over the $y$ coordinate in our numerical model. To account for instrumental
effects, we must also average in the $x$ direction over a distance $\Delta x = D_0 \Delta
\theta$, where $\Delta \theta$ is the angular resolution of the instrument and $D_0$ is
the Sun-Earth distance.
\begin{figure}
\centering
\includegraphics[width=6in]{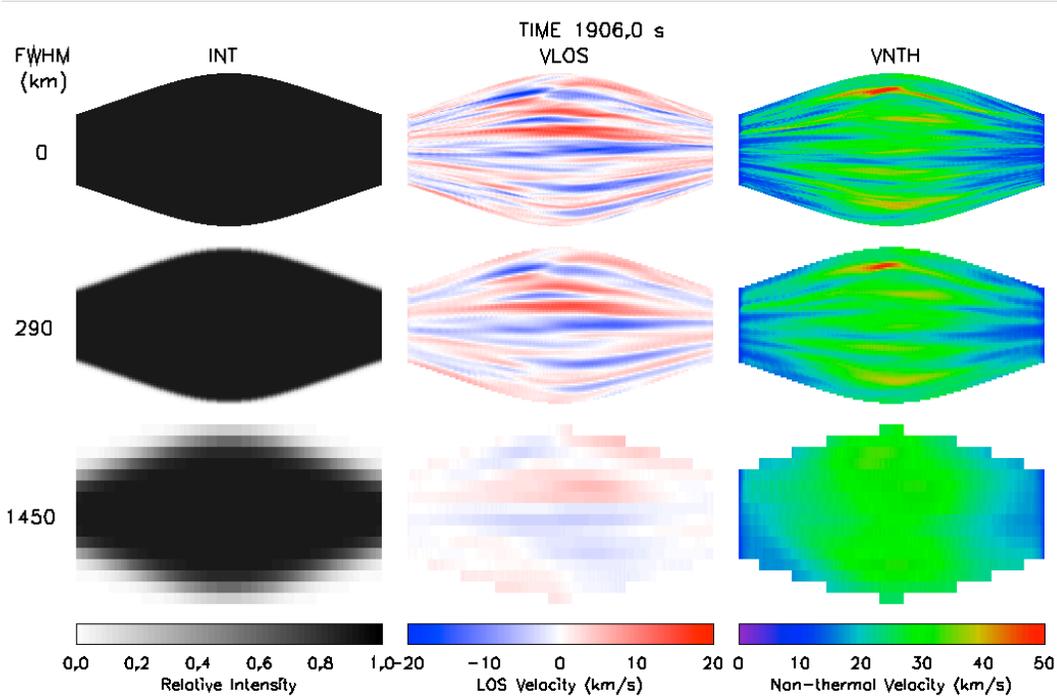}
\caption{Effect of the modeled waves on the Doppler shift and Doppler width of a spectral
line, assuming the loop is viewed from the side. The three column show the intensity (INT),
line-of-sight velocity (VLOS), and non-thermal velocity (VNTH). The three rows correspond to
different values of the spatial resolution of the instrument ($\Delta x =$ 0, 290 and 1450 km).
Each image shows a side view of the loop. The loop curvature is neglected, so the loop axis
runs horizontally through the middle of each image. The transverse (vertical) scale is greatly
expanded compared to the longitudinal scale. The velocity amplitudes are indicated by the
color bars.}
\label{fig8}
\end{figure}

Figure~\ref{fig8} shows the Doppler velocities at time $t = 1906$~s in the simulation.
The three columns show maps of emission intensity (INT), average line-of-sight velocity
(VLOS), and average non-thermal velocity (VNTH). The top row
shows such maps for the full resolution of our numerical code, and the other rows give
maps where spatial smearing and pixelation have been applied to the data. The middle
row represents a hypothetical future instrument with a spatial resolution element size of
290~km (FWHM) and pixel size of 121~km, when projected onto the Sun. The bottom row
is for the EIS instrument on {\it Hinode}, which has a spatial resolution element of about 
1450 km (2") and pixel size of 725 km (1"). In these images only the merged magnetic field is
shown, and the loop curvature is neglected, so each image is the projection of the loop on
the $(s,x)$ plane. Note that the width $w_m (s)$ of the merged field varies with position $s$
along the loop. The vertical size of each image corresponds to a width of 10 Mm, and
the vertical scale is expanded compared to the horizontal scale to show more clearly the
velocity structures inside the loop. The left column shows the intensity using an inverted
grey-scale (black is bright). At full resolution the edges of the observed structure are sharp
because the loop is assumed to have a square cross-section (see section \ref{multi}) and
the emissivity is assumed to vanish outside the simulation domain. At lower resolution the
edges become more fuzzy and the pixel size becomes more obvious.

The middle column of Figure~\ref{fig8} shows the predicted VLOS as a color-scale image.
The velocity scale is given at the bottom of the column. The upper panel
shows the velocity map at full resolution. Note that VLOS changes rapidly in the
$x$ direction (vertical) but gradually in the $s$ direction (horizontal), indicating the internal
motions are coherent along the loop. At the loop top (middle of image) the perpendicular
variations occur on a length scale of about 500 km. The rms value of VLOS over the
entire image is 5.8 $\rm km ~ s^{-1}$. The middle panel shows VLOS as observed
with an instrument that has a spatial resolution of 290~km. The pattern is basically unchanged
(rms velocity 5.1 $\rm km ~ s^{-1}$), indicating that the velocity variations would be well
resolved in such observations. The bottom panel shows the velocity pattern expected for
EIS observations. In this case the velocity amplitude is only 2.7 $\rm km ~ s^{-1}$, which
is below the sensitivity of the EIS instrument. Therefore, EIS is not expected to be able to
observe the VLOS fluctuations simulated here, and indeed, EIS observations do not
show clear evidence for such fluctuations.

The right column of Figure~\ref{fig8} shows the predicted non-thermal velocity (VNTH), and
the corresponding velocity scale is given at the bottom of this column. The value of VNTH
averaged over the image is about 27 $\rm km ~ s^{-1}$, nearly independent of resolution.
This is somewhat higher than the non-thermal velocity of $17.6 \pm 5.3$ $\rm km ~ s^{-1}$
measured with EIS \citep[][]{Brooks2016}, the most-probable velocity of 15 $\rm km ~ s^{-1}$
observed with IRIS \citep[][]{Testa2016}, and the values of 14 -- 26 $\rm km ~ s^{-1}$
observed at the Norikura Solar Observatory \citep[][]{Hara1999}. Therefore, the transverse
waves in our model have relatively high velocities that are only marginally consistent with
the available spectroscopic observations.

\subsection{Summary}
\label{summary}

We have seen that the simulated AWT can produce only enough heat to maintain the coronal
loop at a pressure $p_{\rm cor} = 1.8$ $\rm dyne ~ cm^{-2}$ and peak temperature
$T_{\rm max} = 2.5$~MK. The heating rate $Q$ varies strongly in space and time, but the
time-averaged heating rate in the corona is nearly constant across the loop, and the temperature
fluctuations are predicted to be relatively small (see section~\ref{variation}). In contrast, the
observations show that active regions have broad temperature distributions \citep[e.g.,][]
{Winebarger2011, Warren2011, Warren2012} with peaks of the observed DEM distributions
at temperatures of 3 -- 4 MK. Therefore, the value of $T_{\rm max}$ predicted by our model is
not quite high enough, and the model cannot readily explain the observed broad temperature
distributions of active regions. However, the transverse velocities in our model are already quite
high (see section \ref{doppler}), so enhancing the heating rate by increasing the footpoint velocity
would likely produce a disagreement with the spectroscopic observations. Therefore, it makes
sense to consider an alternative model for coronal heating.

\section{MAGNETIC BRAIDING MODEL}
\label{without_tubes}

The model presented in sections \ref{multi} and \ref{with_tubes} assumes that the flux tubes
are located on a square lattice (see Figure~\ref{fig1}), and the tubes do not move horizontally.
This allowed us to focus on waves {\it inside} the flux tubes, which have periods less than
about 1 minute and are significantly amplified as they propagate upward through the photosphere
and chromosphere. We did not include the longer-term motions of the flux tubes themselves.
However, observations show that  photospheric flux elements exhibit random motions
on a wide range of time scales \citep[e.g.,][] {DeVore1985, Wang1988, Schrijver1990,
Muller1994, Komm1995, Schrijver1996, Berger1996, Hagenaar1999, Utz2010,
Abramenko2011, Wedemeyer2012, Chitta2012}. In the magnetic braiding model the coronal
field lines are assumed to be braided by random footpoint motions on time scales of tens
of minutes to hours, which must be motions of the flux tubes themselves. The observed
random motions can be characterized in terms of a photospheric diffusion constant $D$,
and in magnetic network and plage regions $D$ is in the range 60 -- 250 $\rm km^2 ~ s^{-1}$
for motions on time scales of tens of minutes to a few days \citep[][]{Berger1998}.
It is desirable to include the observed random motions also in our wave heating model.
However, in reduced MHD the magnetic field strength $B_0$ must be constant across the
modeled structures, so we must separately describe each flux tube, as well as the merged field,
and the field-free regions between the flux tubes cannot be included in the model. When the
flux tubes have variable positions in the $(x,y)$ plane, their cross-sections cannot be square,
and their relationship to the merged field becomes too complex to describe in a reduced
MHD model.
\begin{figure}
\centering
\includegraphics[width=6in]{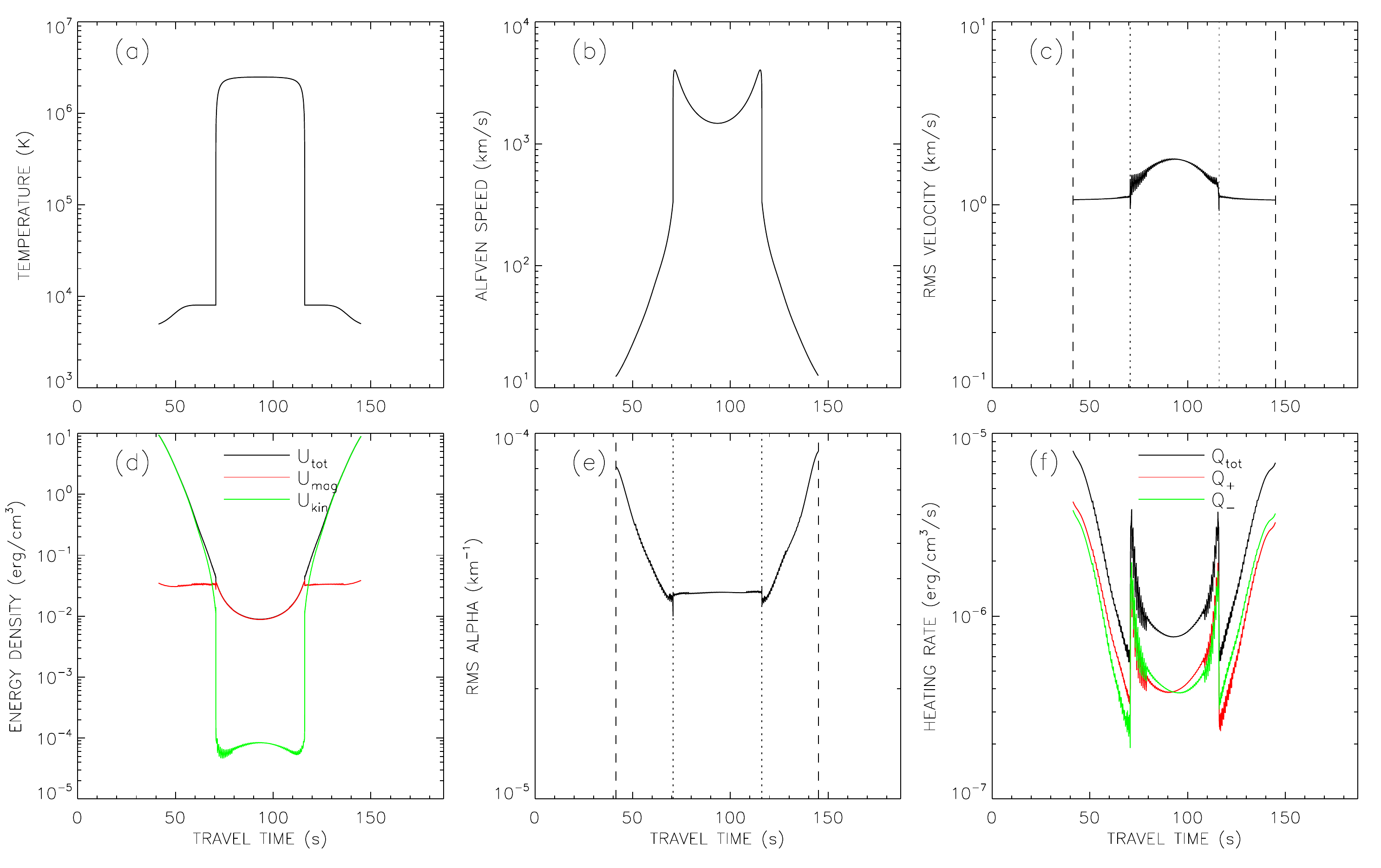}
\caption{Various quantities as function of position along the loop for the magnetic braiding
model: (a) temperature $T$, (b) Alfv\'{e}n speed $v_{\rm A}$, (c) velocity $v_{\rm rms}$,
(d) total energy density $U_{\rm tot}$ and contributions from magnetic- and kinetic energy,
(e) magnetic twist parameter $\alpha_{\rm rms}$, (f) total energy dissipation rate
$Q_{\rm tot}$ (black curve), and contributions $Q_{\pm}$ from the two different wave types
(red and green curves). In panels (c) and (e), the dashed and dotted lines indicate the
merging heights and TRs, respectively.}
\label{fig9}
\end{figure}

In this section we present a model in which the photospheric flux tubes have been removed,
and the ``footpoint" motions are applied directly at the merging heights ($z_m \approx 520$~km).
For each end of the coronal loop the imposed velocity field has components $v_x (x,y,t)$
and $v_y (x,y,t)$, which describe the slow, random motions that would have been produced at
the merging heights by the underlying flux tubes, if they had been included in the model. We find
that such motions produce braiding of the coronal field lines, so we call this the magnetic braiding
model. Note that a variety of ``braiding models" are discussed in the
literature, and different authors use different assumptions for the initial- and boundary
conditions on the braided field. For example, in some models the magnetic field is approximated as
a collection of discrete strands, so the field is not space-filling \citep[e.g.,][]{Berger2009, Berger2015}.
In other models the magnetic field is space-filling, but the photospheric field is concentrated in
discrete flux elements that move about on the solar surface \citep[e.g.,][]{Parker1983, Berger1991,
Berger1993, Priest2002, Janse2010}. It is found that thin current sheets develop at the interfaces
between the flux tubes in the corona, and magnetic reconnection events (``nanoflares") are
predicted to occur in these sheets \citep[][]{Parker1988}. These magnetic braiding models
implicitly assume that the photospheric flux tubes have long lifetimes, and that the flux tubes
can be wrapped around each other for many hours before reconnection is triggered \citep[][]
{Parker1983}. In the present work we assume that the braided field is space-filling, but the
photospheric flux tubes are not included in the model. The velocities $v_x (x,y,t)$ and $v_y (x,y,t)$
at the merging height are assumed to be continuous functions of position, so there are no
topologically distinct flux tubes in the corona. This assumption of spatial continuity of the
footpoint motions is often used in numerical
simulations of the build-up and evolution of braided magnetic fields \citep[e.g.,][] {vanB1988,
Mikic1989, Galsgaard1996, Rappazzo2007, Rappazzo2008, Rappazzo2013, Dahlburg2016}.
Other authors focus on the turbulent relaxation of braided fields, and assume that the
braided field is present in the initial state of the simulation \citep[e.g.,][]{Wilmot-Smith2015,
Pontin2015, Pontin2017}, so in these models the build-up of the braided field is not simulated.
For the build-up of a braided field to occur, the onset of reconnection must be delayed as
long as possible \citep[][]{Pontin2015}. It is still an open question whether the observed random
motions of photospheric magnetic elements can produce the kind of braided fields needed in
nanoflares models.

In the present model we assume that $v_x (x,y,t)$ and $v_y (x,y,t)$ are continuous functions
of position, so we neglect the spatial discontinuities in velocity that would naturally occur at the
boundaries between neighboring flux tubes at the merging height. This has the effect of delaying
the onset of reconnection as long as possible, and maximizes the build-up of energy in the braided
field.
Observations suggest that the motions of neighboring flux elements are uncorrelated. Therefore,
we assume that the correlation length $L_\perp$ of the velocity field is equal to the width of a flux
tube at the merging height, $L_\perp \approx$ 1,100 km, and there are 28 driver modes.
The rms velocity is $v_{\rm rms} = 1$ $\rm km ~ s^{-1}$, and the velocity auto-correlation time
is $\tau_{\rm c} = \tau_0 / \sqrt{2 \pi} \approx 399$~s, where $\tau_0 = 1000$~s is the Fourier
filtering parameter described in paper~I. Using these values, we determine the random velocity
field at each end of the loop. We then select 300 points randomly distributed over the simulation
domain at the merging height, and follow these points for 6000~s as they move with the flow.
The mean square displacement $< r^2 (t) >$ of the particles increases more or less linearly with
time, as expected for random walk. By fitting the results to the expression $< r^2 (t) > = 4 D t$,
we find $D \approx 200$ $\rm km^2 ~ s^{-1}$, which is at the high end of the range of
observed diffusion constants \citep[see][]{Berger1998}.

The structure of the background field in the magnetic braiding model is exactly the same
as in the AWT model, and the methods for solving the reduced
MHD equations are the same as in section \ref{wave} for the merged field. Since the flux
tubes are omitted, we cannot simulate the short-period waves that emanate from these flux
tubes. However, we assume that the heating produced by such short-period waves is still
present. Therefore, we take $p_{\rm cor} = 1.8$ $\rm dyne ~ cm^{-2}$, the same as in the
AWT model (section \ref{with_tubes}). Hence, the background atmosphere for the magnetic
braiding model is the same as that shown for the merged field in Figure~\ref{fig2}.

The dynamics of the braided field was simulated for a period of 8000~s.
Figure~\ref{fig9} shows various quantities as function of position along the loop,
expressed in terms of the wave travel time $t_0 (s)$. As a reference, Figures~\ref{fig9}(a)
and \ref{fig9}(b) show the temperature $T(s)$ and Alfv\'{e}n speed $v_{\rm A} (s)$, which
are the same as in Figure~\ref{fig2}. The other panels show various quantities averaged
over the loop cross-section and over time. Figure~\ref{fig9}(c) shows the rms velocity of the
waves. Note that the peak velocity in the corona is only about 1.8 $\rm km ~ s^{-1}$.
Figure~\ref{fig9}(d) shows the total energy density $U_{\rm tot}$
of the perturbations (black curve), the magnetic energy density $U_{\rm mag}$ (red curve),
and the kinetic energy density $U_{\rm kin}$ (green curve). In the corona $U_{\rm mag} \gg
U_{\rm kin}$, so the perturbations are dominated by the magnetic field, as expected for
a magnetic braiding model. However, in the chromosphere the kinetic energy dominates,
even though the rms velocity is only about 1.1 $\rm km ~ s^{-1}$. Figure~\ref{fig9}(e) shows
the rms value of the twist parameter, $\alpha \equiv ( \nabla \times {\bf B} )_\parallel / | {\bf B} |$.
Note that $\alpha_{\rm rms} (s)$ is nearly constant in the coronal part of the loop, indicating
that the magnetic field is close to a nonlinear force-free state, $\nabla \times {\bf B} \approx
\alpha {\bf B}$, with $\alpha$ constant along the field lines. However, $\alpha_{\rm rms} (s)$
is not constant in the chromospheres at either end of the loop, which is due to wave-like
perturbations in the lower atmosphere.

\begin{figure}
\centering
\includegraphics[width=6in]{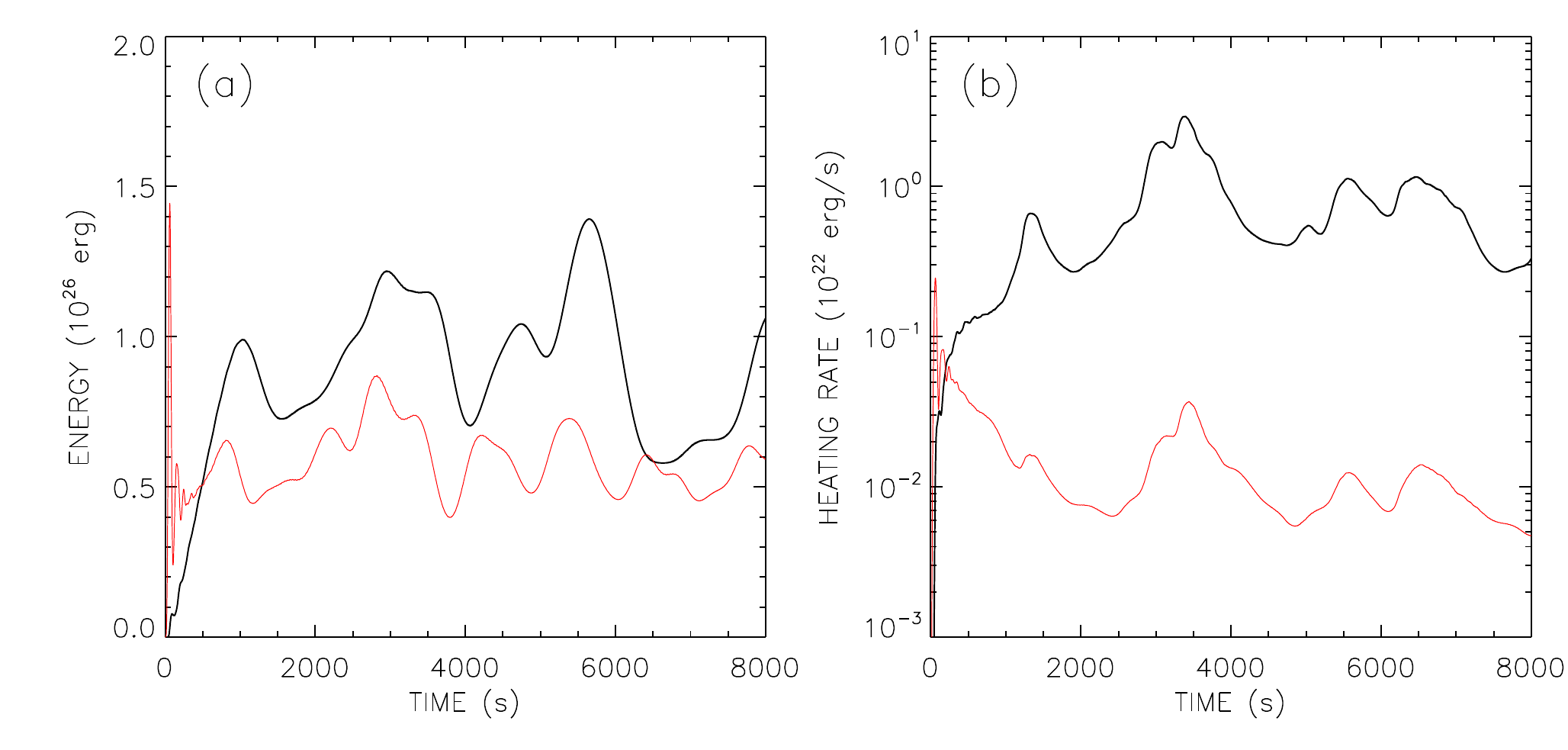}
\caption{Free energies and heating rates as functions of time for the magnetic braiding
model. (a) Total energy $W_{\rm cor} (t)$ integrated over the coronal part of the loop
(black curve), which is dominated by magnetic free energy of the braided field, and total
energy $W_{\rm chrom} (t)$ integrated over the chromospheric parts of the loop (red
curve), which is dominated by kinetic energy. (b) Total heating rates integrated over the
coronal and chromospheric parts of the loop (black and red curves, respectively).}
\label{fig10}
\end{figure}

Figure~\ref{fig9}(f) shows the total energy dissipation rate $Q_{\rm tot} (s)$ (black curve)
for the magnetic braiding model, as well as the contributions $Q_{\pm}$ from perturbations
propagating in opposite directions along the loop. Comparison with Figure~\ref{fig3}(d)
shows that these dissipation rates are much smaller than those in the AWT model.
For example, in the corona $Q_{\rm tot} \sim 10^{-6}$ $\rm erg ~ cm^{-3} ~ s^{-1}$,
whereas in the AWT model $Q_{\rm tot} \sim10^{-3}$ $\rm erg ~ cm^{-3} ~ s^{-1}$.
Therefore, the magnetic braids associated with the slow random motions of the flux tubes
add an insignificant amount of heat compared to the energy provided by the short-period
waves inside the flux tubes. By itself, the contribution from magnetic braiding
is much smaller than the heating rate needed to maintain the background atmosphere
with $T_{\rm max} = 2.5$ MK and $p_{\rm cor} = 1.8$ $\rm dyne ~ cm^{-2}$. The main
reason for the low heating rates compared to the AWT model is that there is almost no
amplification of the waves as they propagate upward through the chromosphere.
The rms velocity in the corona is only about 1.1 -- 1.8 $\rm km ~ s^{-1}$, much
smaller than the velocity in the AWT model. This lack of wave amplification is due to the
long correlation time $\tau_{\rm c}$ of the footpoint motions compared to the wave travel
time in the chromosphere.

The total energy $W_{\rm cor} (t)$ of the disturbances in the coronal part over the loop was
computed by integrating the energy density $U_{\rm tot} (s,t)$ over the coronal part of the
volume, including the TRs at the two ends. Similarly, the energy $W_{\rm chrom} (t)$ in the
chromosphere was determined by integrating over the two chromospheric parts of the
volume. Figure~\ref{fig10}(a) shows $W_{\rm cor} (t)$ (black curve) and $W_{\rm chrom} (t)$
(red curve), plotted as functions of time $t$ in the simulation. The coronal energy increases
with time over the first 1000 seconds of the simulation, but then reaches a quasi-steady state in
which the energy fluctuates around a mean value. Note that $W_{\rm chrom}$ is comparable
to $W_{\rm cor}$, even though the chromosphere represent only a small fraction of the loop
volume. The coronal energy $W_{\rm cor} (t)$ is dominated by magnetic free energy, while
$W_{\rm chrom} (t)$ is dominated by the kinetic energy of the plasma  (see
Figure~\ref{fig9}(d)). Figure~\ref{fig10}(b) shows similar plots for the volume
integrated heating rates in the corona (black curve) and in the chromosphere (red curve).
This figure demonstrates that most of the injected energy is dissipated in the corona, as
expected for a magnetic braiding model. The main problem with this model is that the
amount of energy involved is small compared to the radiative and conductive losses.

The above results may be compared with other magnetic braiding models. For example,
\citet[][]{Dahlburg2016} simulated magnetic braiding in a coronal loop, using a full MHD
description of the plasma. The computational domain extends from the base of the TR at
one end of the loop to the TR at the other end, and random footpoint motions with velocity
$v_{\rm rms} = 1$ $\rm km ~ s^{-1}$ are applied at the TR boundaries. The background
magnetic field is assumed to be uniform, and different values of the magnetic field strength
$B_0$ are considered (100, 200 and 400 G). The footpoint motions produce braided
magnetic fields, and the associated field-aligned electric currents produce Ohmic heating
of the coronal plasma. The authors simulate the response of the coronal plasma to such
heating, and find that the heating produces large variations in coronal  temperature across
the loop, but only small density variations. \citet[][]{Dahlburg2016} compare the simulated
DEM distributions with observations, and find that a good fit to the observations is obtained
for a coronal field strength of 400~G. However, the Dahlburg et al.~model has
several limitations that may cause the coronal heating rates and temperatures to be
overestimated. First, the model does not include the chromosphere, which may cause the
energy available to the corona to be overestimated. Second, the model does not take into
account coronal loop expansion, and assumes a magnetic field strength of 400~G, which
is high compared to the values typically found in extrapolations of photospheric
magnetograms. For example, in our previous work on observed coronal loops in the cores
of active regions we obtained minimum field strengths in the range 13 -- 30 G (paper~II),
30 -- 120 G (paper~III), and 20 -- 45 G (paper~IV). Finally, in the model by \citet[][]
{Dahlburg2016} the temperature and density are held fixed at the TR boundaries,
so there is no chromospheric ``evaporation" in response to coronal heating events.
Therefore, the model may overestimate the temperature increase resulting from a given
heating event.

\section{Discussion}

In this paper we developed a model for the heating of a magnetic loop containing multiple
flux tubes in the photosphere. The flux tubes merge at a height $z_m \approx 520$ km in
the upper photosphere, so the magnetic field in the chromosphere and corona is a collection
of flux tubes that are pressed up against each other. We simulated the dynamics of Alfv\'{e}n
waves in such a magnetic structure, using a reduced MHD model. The waves are assumed
to be produced by random footpoint motions {\it inside} the photospheric flux tubes.
The turbulent waves were simulated for a period of 3000~s, comparable to the loop cooling
time (about 2000 s). We found that AWT can produce enough heat to maintain a coronal
pressure $p_{\rm cor} = 1.8$ $\rm dyne ~ cm^{-2}$ and peak temperature $T_{\rm max} =
2.5$~MK. The wave heating rate $Q$ varies strongly in space and time, but the time-averaged
rate $\overline{Q}$ in the corona is nearly constant over the loop cross-section, so the model
does not predict large temperature or density variations across the loop.
In contrast, the observed DEM distributions indicate that active regions have broad temperature
distributions with peak temperatures of 3 -- 4 MK but extending to significantly higher and
lower temperatures \citep[e.g.,][]{Winebarger2011, Warren2011, Warren2012}. Therefore,
the value of $T_{\rm max}$ predicted by our AWT model is not quite high enough, and the
model cannot readily explain the observed broad DEM distributions.

We simulate the effects of AWT on the Doppler shift and Doppler broadening of spectral
lines formed in the corona (section \ref{doppler}). We predict that the Alfv\'{e}n waves should
be detectable as variations in Doppler shift, provided the instrument has sufficiently high
spatial resolution (FWHM $< 0.5"$). The rms value of the non-thermal velocity is predicted
to be nearly independent of spatial resolution, and is about 27 $\rm km ~ s^{-1}$, which
is high compared to the observed values \citep[][]{Brooks2016, Testa2016, Hara1999}.
Therefore, despite not providing enough heating, the AWT model is already injecting as
much energy into the corona as is consistent with spectroscopic observations.

The heating rates $Q(t)$ for co-moving points vary strongly with time, and can be described
as a series of heating events. The integrated energy $\int Q(t) dt$ of an event can be compared
with the thermal energy density of the plasma. In the chromosphere the bursts have enough
energy to significantly increase the temperature, which may be important for understanding
type~II spicules and other dynamic phenomena in the chromosphere \citep[][]
{DePontieu2007}. However, in the corona the bursts have only modest energy compared to
the thermal energy density of the coronal plasma, and the temperature fluctuations
produced by such bursts are predicted to be relatively small (see section~\ref{variation}).
The observations indicate there are large changes in temperature as the loops heat up and
cool \citep[e.g.,][]{Viall2011,Viall2012}. Our model cannot directly explain such observations.

The model predicts that the time-averaged heating rate $\overline{Q}$ is enhanced at the
boundaries between the flux tubes in the chromosphere (these boundaries are separatrix
surfaces in our model). In the region
just above the merging height the spatially averaged heating rate $Q_{\rm tot}$ is enhanced
by about a factor 2 compared to its value just below the merging height. However, this extra
heating at the flux tube boundaries does not extend to large heights. The reason for this
rapid decline with height of the boundary heating is that the waves in the chromosphere
are quite turbulent, and produce large displacements of the boundary surfaces. Therefore,
the extra energy associated with the discontinuities in the velocity field at the merging height
is quickly dissipated.

In our first model (section \ref{with_tubes}) we assumed that the photospheric flux tubes have
fixed positions,
so we did not take into account the observed random motions of magnetic flux elements
at the solar surface \citep[e.g.,][]{Wang1988, Schrijver1990, Muller1994, Schrijver1996,
Berger1998, Hagenaar1999, Abramenko2011}. To estimate the effects of such random
motions on coronal heating, we presented (in section \ref{without_tubes}) a second model
in which the flux tubes are omitted and the footpoint motions applied directly at the merging
heights. The imposed random motions have a correlation time $\tau_c \approx 400$ s, and
have a random walk diffusion constant of about 200 $\rm km^2 ~ s^{-1}$, which is
at the high end of the observed range of photospheric diffusion constants
\citep[see][and references therein]{Berger1998}. In this ``magnetic braiding" model the
footpoint motions produce quasi-static braiding of the coronal field lines. We find that
the model without photospheric flux tubes and long correlation time produces much less
coronal heating than the AWT model with flux tubes. We conclude that the observed random
motions on times scales longer than a few minutes are not the main source of energy for
coronal heating in active regions. The short-period waves inside the flux tubes appear
to be more important for the energy balance of the coronal plasma.

The models described in this paper lack the kind of impulsivity usually required to explain
the observed characteristics of the coronal plasma. To have more impulsive heating would
require that magnetic free energy builds up in the corona over a period of several hours before
the energy is released by a reconnection event \citep[e.g.,][]{Parker1983, Pontin2015}. However,
we find that such energy build-up does not occur in either of our models. In the AWT model the
counter-propagating waves interact so strongly that the wave energy is dissipated on a time
scale of only about 100~s, too small for significant energy build-up to occur. In the magnetic braiding
model the dissipation time is much longer ($\sim 10^4$~s), but we find that the amount of energy
released is simply too small to explain the observed coronal temperatures. Therefore, we do
not find the conditions necessary to produce strong heating events. In nanoflare heating models
\citep[see reviews by][]{Klimchuk2010, Cargill2015}, the magnitude of a coronal heating event
is a free parameter that can be adjusted to obtain agreement with solar observations.
However, in our simulation study only the footpoint motions can be imposed, and the magnitude
of heating events cannot be adjusted. Our models use realistic coronal field strengths and
photospheric footpoint motions, so we believe our simulation results put important constraints
on the magnitudes of heating events that can be produced by the observed footpoint motions.

As mentioned above, the AWT model described in section \ref{with_tubes} cannot readily
explain the observed broad DEM distributions of active regions. In the remainder of this section
we discuss several possibilities for resolving this problem. First, the present model
may underestimate the amount of energy injected into the coronal part of the loop.
The flux tubes are modeled using the zeroth order thin-tube approximation, which is not very
accurate at larger heights in the photosphere \citep[][]{YellesChaouche2009}. The merging
of flux tubes is described in a simple way (see Figure~\ref{fig1}), which does not accurately
represent the complex 3D magnetic structure of a plage region \citep[][]{Buehler2015}.
Potential-field modeling suggests that the transverse velocities in the chromosphere may be
enhanced by the presence of magnetic nulls at the merging height \citep[][]{vanB1998,
vanB2003}, but the present model  does not include such null points.

Another limitation of the present modeling is that it is based on the so-called reduced MHD
approximation, so it does not take into account flows along the background magnetic field
or density  variation across the field lines. It is well known that in the upper chromosphere
and TR there are strong flows along the field lines in the form of jets and spicules
\citep[e.g.][] {Beckers1972, Sterling2000, DePontieu2007, Tian2014}, and such flows may
in fact be caused by localized heating and/or wave pressure forces in the chromosphere
\citep[e.g.,][]{Murawski2015, Cranmer2015b}. Therefore, some of the wave energy may be
channeled into parallel flows, which is not taken into account in the present model.
The jet-like flows cause the TR to be highly corrugated, with large variations in density
across the field lines \citep[][]{Feldman1997, Zhang1998, Peter2013}.
Therefore, the present version of the AWT model may overestimate the degree of wave
reflection at the TR, and underestimate the amount of energy transmitted into the corona.
To address this problem full 3D MHD simulations of coronal loops and their connections
to the lower atmosphere are needed. Such simulations are very challenging because of the
high spatial and temporal resolution required in the corrugated TRs.

The strength of the magnetic disturbances at the footpoints of the hot loops may have been
underestimated. During the growth phase of active regions, new magnetic flux emerges
into the solar atmosphere from the convection zone below, and the newly emerged field is
generally not aligned with the preexisting field \citep[e.g.,][]{Ortiz2014}. The mis-alignment
angles may be 10 -- 20 degrees, significantly larger than the values predicted by the models
presented in this paper. Magnetic reconnection between the old and new flux
systems could produce some or all of the observed high temperature emission. The role of flux
emergence in producing the observed DEM distributions should be further investigated.
Also, some hot loops may be anchored in sunspot penumbrae, which are known to have
different orientation angles of the magnetic fields in bright and dark penumbral filaments
\citep[see review by][]{Borrero2011}. These complex penumbral fields could cause more
magnetic energy to be injected into the corona. However, not all hot loops are connected to
sunspot penumbrae.

Finally, the model may overestimate the rate of cooling of the coronal plasma. In standard loop
models the thermal conductivity is assumed to be dominated by Coulomb collisions, and is
given by Spitzer's formula \citep[][] {Cohen1950, Spitzer1953, Spitzer1962}. The effects of wave
turbulence on pitch-angle scattering of non-thermal and thermal electrons have recently been
considered in the context of solar flares to explain the observed rates of temperature decrease
in post-flare loops \citep[][]{Kontar2014, Bian2016a, Bian2016b}. We suggest that similar effects
of pitch-angle scattering may occur in non-flaring loops, producing a reduction in the thermal
conductivity compared to Spitzer's value. In preliminary modeling using different values of the
conductivity parameter $\kappa_0$, we have found that reducing this parameter by a factor 3 -- 5
allows the AWT model to reproduce the observed temperatures in the range 3 -- 4 MK. However,
a simple reduction of $\kappa_0$ may not be the best way to describe the effects of plasma
turbulence. Such a reduction of the thermal conductivity would also affect the thermal stability
of coronal loops \citep[e.g.,][]{Mikic2013, Lionello2013}, and may play a role in the formation of
coronal rain \citep[][] {Antolin2012, Antolin2015}.

\acknowledgements
We thank the referee for his/her comments on the paper.
CHIANTI is a collaborative project involving George Mason University, the University of
Michigan (USA) and the University of Cambridge (UK).
This project was supported under contract NNM07AB07C from NASA to the Smithsonian
Astrophysical Observatory (SAO) and contract SP02H1701R from Lockheed Martin Space
and Astrophysics Laboratory (LMSAL) to SAO.

\appendix

\section{Numerical Methods}

The numerical methods used in the present work are similar to those used in simulating
Alfv\'{e}n waves in the solar wind (see paper~V). The waves are described using stream
functions $f_\pm (x,y,s,t)$ for the Elsasser variables. The dependence of these functions
on the perpendicular $x$ and $y$ coordinates are described as follows:
\begin{equation}
f_{\pm} (x,y,s,t) = \sum_{k=1}^M f_{\pm,k} (s,t) F_k (\tilde{x},\tilde{y}) ,  \label{eq:fpm}
\end{equation}
where $\tilde{x}$ and $\tilde{y}$ are dimensionless perpendicular coordinates,
$F_k (\tilde{x},\tilde{y})$ is an eigenfunction of the $\nabla_\perp^2$ operator,
$k$ is an index enumerating the different eigenfunctions, $M$ is the number of
eigenmodes, and $ f_{\pm,k} (s,t)$ are the mode amplitudes describing the waves.
Two different sets of eigenfunctions are used, one for the
flux tubes and another for the merged field. For the flux tubes we use $\tilde{x} \equiv
x/R(s) + 0.5 $ and $\tilde{y} \equiv y/R(s) + 0.5 $, where $R(s)$ equals the width of
a flux tube, $R(s) = w_t (s)$, so the dimensionless coordinates are in the range
$0 \le \tilde{x} \le 1$ and $0 \le \tilde{y} \le 1$. The eigenfunctions satisfy ``closed''
boundary conditions with $F_k = 0$ at the side boundaries of the tubes, so they are
products of sine functions:
\begin{equation}
F_k (\tilde{x},\tilde{y}) \equiv F_{n_x,n_y} (\tilde{x},\tilde{y}) =
2 \sin (n_x \pi \tilde{x}) \sin (n_y \pi \tilde{y}) .  \label{eq:Fk}
\end{equation}
For flux tubes $n_x$ and $n_y$ are integers in the range 1 to 42, so there are $M = 1764$
eigenmodes, and equation (\ref{eq:fpm}) is equivalent to a 2-dimensional (2D)
sine transform.
For the merged field we use $\tilde{x} \equiv x/R(s)$ and $\tilde{y} \equiv y/R(s)$, where
$R(s) = w_m (s)/2$ is the half-width of the computational domain, so the dimensionless
coordinates  are in the range $0 \le \tilde{x} \le 2$ and $0 \le \tilde{y} \le 2$. Also, we use
periodic boundary conditions, so the eigenfunctions are products of sine and cosine functions,
and equation (\ref{eq:fpm}) is now equivalent to an ordinary 2D Fourier transform (for the
merged field $M =$ 29,240). Each eigenmode has a well-defined dimensionless wavenumber
$a_k$ (in units of $1/R$), and the maximum wavenumber in one direction is $a_{\rm max}  =
42 \pi$ for flux tubes and $a_{\rm max}  = 85 \pi$ for the merged field.

Inserting equation (\ref{eq:fpm}) into the reduced MHD equations (\ref{eq:rmhd1}), we find
\begin{eqnarray}
\frac{\partial \omega_{\pm,k}} {\partial t} & = & \mp v_{\rm A}  \frac{\partial \omega_{\pm,k} }
{\partial s}  + \frac{1}{2} \frac{dv_{\rm A}} {ds}  \left( \omega_{+,k} - \omega_{-,k} \right)
\nonumber \\
 & & + \frac{1}{2 R^4} \sum_j \sum_i M_{kji} ( a_i^2 - a_j^2 - a_k^2 ) f_{\pm,j} f_{\mp,i}
- \nu_{\pm,k} ~ \omega_{\pm,k} , \label{eq:rmhd2}
\end{eqnarray}
where $\omega_{\pm,k} (s,t)$ is the wave vorticity for each mode. The four terms on the
right-hand side describe wave propagation, linear couplings (including wave reflection),
nonlinear couplings, and wave damping. Here $M_{kji}$ is a dimensionless coupling matrix
(see equation (5) in paper~V), and the damping rates $\nu_{\pm,k} (s,t)$ are discussed below.
The nonlinear terms are evaluated using Fourier Transform
methods. For each position $s$ along the loop, the quantities inside the bracket operators
in equation (\ref{eq:rmhd1}) are computed by transforming these quantities from the spectral
to the spatial domain, where the brackets are easily evaluated. Then the results are
transformed back to the spectral domain to determine the nonlinear term in equation
(\ref{eq:rmhd2}). For the merged field the transformation involves a spatial grid of
$256 \times 256$ points. For flux tubes the sine transform is performed as an ordinary
Fourier transform on a grid twice the width of one flux tube ($128 \times 128$ grid),
and the calculation is done separately for each flux tube.

The coupled equations (\ref{eq:rmhd2}) are solved by using finite differences in the coordinate
$s$ along the background field. To accurately describe the wave reflections at the TRs,
we use a highly non-uniform grid, $s_n$, where $n$ is an index of the grid points. In the TRs
the steps $\Delta s_n$ between neighboring grid points must be very small (less than 1 km),
which means that the Alfv\'{e}n travel time $\Delta t_n = \Delta s_n / v_{A,n} $ between
grid points is also very small (less than 1 ms). Therefore, to properly simulate the
dynamics of the waves it is necessary to take very small time steps at those
heights. Elsewhere along the loop the variations in the background atmosphere are
more gradual and we can take larger time steps (up to 1 s). We developed a method
for dealing with different time steps in different regions along the loop. Let $\Delta t_0$ be the
maximum time step to be used (for the models presented in this paper $\Delta t_0 = 0.811$~s).
The grid $s_n$ is chosen such that the Alfv\'{e}n travel time $\Delta t_n$ is constant within
certain sections of the grid, and jumps by a factor 2 between sections. Therefore,
within each section $\Delta t_n = \Delta t_0 2^{-N}$, where $N$ is the level of
refinement for that section ($0 \le N \le 14$). Furthermore, within each section
we use a time step $\Delta t$ equal to the wave  travel time $\Delta t_n$ between
grid points. Let $\omega_{\pm,n}$ and $\omega_{\pm,n+1}$ be the vorticities of the
waves at grid points $n$ and $n+1$ at time $t$, where we omit the dependence
on mode index $k$. Then the vorticities at time $t+\Delta t$ are given by
\begin{eqnarray}
\omega_{+,n+1}^\prime & = & ( 1 + \epsilon_n ) \omega_{+,n}
- \epsilon_n \omega_{-,n+1} ,  \label{eq:omegp} \\
\omega_{-,n}^\prime ~~~ & = & \epsilon_n \omega_{+,n}
+ (1 - \epsilon_n ) \omega_{-,n+1} , \label{eq:omegm}
\end{eqnarray}
where $\epsilon_n \equiv (v_{A,n+1} - v_{A,n} ) / (v_{A,n+1} + v_{A,n} )$ is related
to the gradient in Alfv\'{e}n speed. To evolve the waves over a full time step
$\Delta t_0$ the sections of the grid with refinement level $N$ must be processed
$2^N$ times, so highly refined sections are processed much more frequently.
The order in which the different sections are processed is such that all sections evolve
more or less equally in time, with no more than one time step $\Delta t_n$ between
neighboring sections. The nonlinear and damping terms in the reduced MHD equations
are evaluated only at the full time steps $\Delta t_0$. We find that this technique produces
reasonably accurate results for the propagation and reflection of waves in the TR.

For the AWT model the footpoint motions are imposed at the bottom of the flux tubes,
$s = 0$ and $s = L$. The vorticities $\tilde{\omega}_{k} (t)$ of the three driver modes
are computed as described in section \ref{footpoint}, and for all other modes we set
$\tilde{\omega}_{k} (t) = 0$. The boundary conditions are implemented by setting
the amplitudes of the upward propagating waves as follows:
\begin{equation}
\omega_{+,k} (t) = 2 \tilde{\omega}_{k} (t) - \omega_{-,k} (t)   ~~~~ \hbox{at $s = 0$} ,
\label{eq:BC1}
\end{equation}
and similar for the waves $\omega_{-,k} (t)$ at $s = L$. Equation (\ref{eq:BC1})
implies that the downward propagating waves in the flux tubes are strongly reflected
at the base of the photosphere, but this is not a problem in the AWT model because the
waves are quite turbulent and no standing waves are created. However, for the magnetic
braiding model the footpoint motions are applied at the merging heights, and we find that
equation (\ref{eq:BC1}) leads to the formation of a standing wave that resonates between
the merging height and the TR. This wave has a period $P \approx 70$~s. To suppress
this standing wave, we modify equation (\ref{eq:BC1}) as follows:
\begin{equation}
\omega_{+,k} (t) = 2 \tilde{\omega}_{k} (t) - \onehalf \left[ \omega_{-,k} (t) +
\omega_{-,k} (t- \onehalf P) \right]   ~~~~ \hbox{at $s = z_m$} ,  \label{eq:BC2}
\end{equation}
and similar for the wave $\omega_{-,k} (t)$ at $s = L - z_m$. Equation (\ref{eq:BC2})
implies that a downward propagating wave with period $P$ is transmitted at the
merging height, while longer-period waves are reflected. Therefore, the imposed
footpoint motions, which have a correlation time of about 400~s, are almost unaffected by
this modification of the boundary conditions in the magnetic braiding model.

The term in equation (\ref{eq:rmhd1}) involving the viscosity $\tilde{\nu}_{\pm}$
produces a wave damping rate $\nu_{\pm,k}$ given by
\begin{equation}
\nu_{\pm,k} = \nu_{\pm} \left( \frac{a_k} {a_{\rm max}} \right)^2 ,  \label{eq:nuk}
\end{equation}
where $\nu_{\pm} \equiv \tilde{\nu}_{\pm} ( a_{\rm max} / R)^2$. This rate depends
quadratically on wavenumber, which is different from the ``hyperdiffusion" used in
our earlier work (papers I through IV). The use of ordinary diffusion is made possible
by the fact that we now have much higher spatial resolution in $x$ and $y$ compared to
our earlier models. Most of the
wave damping occurs at high wavenumbers, $a_k > (2/3) a_{\rm max}$, which we call
the dissipation range. The maximum damping rates $\nu_\pm (s,t)$ are chosen such that
the wave damping time is comparable to the cascade time scale. For the AWT model 
(section~\ref{with_tubes}) we set $\nu_{\pm} = 1.5 k_d Z_{\mp,\rm d}$, where $k_{\rm d} =
(2/3) a_{\rm max} / R$ is the wavenumber at the start of the dissipation range, and
$Z_{\mp, \rm d}$ are the Elsasser variables just below this range. The latter are given
by sums over all modes with wavenumbers in the range $(1/2) a_{\rm max} < a_k < (2/3)
a_{\rm max}$ [see equation (23) in paper~V]. The damping rates $\nu_\pm (s,t)$ are
also smoothed in time, using a running time average with time constant of 500~s.
For the magnetic braiding model (section~\ref{without_tubes}) the Elsasser variables are
dominated by magnetic fluctuations, so we use $\nu_{+} = \nu_{-} = 1.5 k_d V_{\rm d}$,
where $V_{\rm d}$ is the total velocity of modes in the above wavenumber range.

The energy dissipation rates averaged over the loop cross-section can be computed by
adding the contributions from all wave modes:
\begin{equation}
Q_{\pm} (s,t) = \frac{\rho}{2} \sum_{k=1}^M \nu_{\pm,k} \left( \frac{a_k}{R} \right)^2
f_{\pm,k}^2 = \frac{\rho \nu_{\pm}} {2 R^2 a_{\rm max}^2} \sum_{k=1}^M a_k^4 f_{\pm,k}^2 ,
\label{eq:Qst}
\end{equation}
and below the merging heights these quantities are further averaged over all flux tubes.
The total energy dissipation rate is $Q_{\rm tot} (s,t) \equiv Q_{+} + Q_{-}$.
Position dependent dissipation rates are computed as follows:
\begin{equation}
Q_\pm (x,y,s,t) = \frac{1}{2} \rho \nu_\pm \left( \frac{R} {a_{\rm max}} \right)^2
 \left( S_{\pm,xx}^2 + S_{\pm,xy}^2 \right) ,
\label{eq:Qxyst}
\end{equation}
where $S_{\pm,xx}$ and $S_{\pm,xy}$ are the diagonal and off-diagonal components
of the rate of strain tensor for the Elsasser variables:
\begin{eqnarray}
S_{\pm,xx} (x,y,s,t) & \equiv & 2 \frac{\partial z_{\pm,x}} {\partial x} =
2 \frac{\partial^2 f_{\pm}} {\partial x \partial y} , \label{eq:Sxx} \\
S_{\pm,xy} (x,y,s,t) & \equiv & \frac{\partial z_{\pm,y}} {\partial x} +
\frac{\partial z_{\pm,x}} {\partial y} =  \frac{\partial^2 f_{\pm}} {\partial y^2}
- \frac{\partial^2 f_{\pm}} {\partial x^2} .  \label{eq:Sxy}
\end{eqnarray}
Using the orthogonality of the eigenmodes, it can be shown that the average of equation
(\ref{eq:Qxyst}) over the loop cross-section equals expression (\ref{eq:Qst}). The position
dependent rates are determined by first computing $S_{\pm,xx}$ and $S_{\pm,xy}$ in the
spectral domain, and then Fourier transforming them to the spatial domain. The total
dissipation rate shown in Figure~\ref{fig6} is given by $Q (x,y,s,t) = Q_{+} + Q_{-}$.

\section{Coronal Loop Model}

The background atmosphere for the reduced MHD model is constructed by solving the energy
balance equations for a coronal loop. We consider a thin magnetic flux tube with length
$L_{\rm c}$ in the corona, as measured from the base of the TR at one end of the loop to the
base of the TR at the other end. The cross-sectional area $A_0 (s)$ of the tube is an arbitrary
function of position $s$ along the loop, and is related to the magnetic field strength $B_0 (s)$
such that $A_0 (s) B_0 (s) = \Phi =$ constant. We assume that the plasma heating is either steady
in time or fluctuates on a time scale much shorter than the loop cooling time (high-frequency
heating). Furthermore, variations in heating rate across the loop are neglected. Then the heating
rate $Q_{\rm A} (s)$ can be taken to be constant in time. The coronal loop is assumed to be
thermally stable, so the temperature $T (s)$ and density $\rho (s)$ are also constant in time.
The model can also describe asymmetric loops, so there may be a steady mass flow along
the loop with velocity $v(s)$. For the purpose of this Appendix the endpoints of the loop are
assumed to be located at $s = 0$ and $s = L_{\rm c}$, where $L_{\rm c}$ is the coronal loop
length. These endpoints are defined as those points in the TR where the temperature equals
twice the chromospheric temperature, $T (0) = T(L_{\rm c}) = 2 T_0$, where $T_0 = 8000$~K.
The gas pressure $p(s)$, mass density $\rho (s)$ and electron density $n_e (s)$ are given
by the ideal gas law:
\begin{equation}
p = c_1 n_p k_B T, ~~~ \rho = c_2 m_p n_p, ~~~ n_e = c_3 n_p ,
\end{equation}
where $n_p (s)$ is the proton density, $m_p$ is the proton mass, and $k_B$ is the
Boltzmann constant. The plasma is assumed to be fully ionized. The constants $c_i$ are
given by $c_1 = 2 + 3 A_{\rm He}$, $c_2 = 1 + 4 A_{\rm He}$, and $c_3 = 1 + 2 A_{\rm He}$, 
where $A_{\rm He} = 0.1$ is the helium abundance.
For simplicity we neglect the effects of gravity in the corona, and we assume that  the flow
velocity is small compared to the sound speed. Then the gas pressure $p(s) = p_{\rm cor}$
= constant. The coronal pressure $p_{\rm cor}$ is treated as a free parameter of the model.
In the following we assume $B_0 (s)$, $L_{\rm c}$ and $p_{\rm cor}$ to be known quantities.

In steady state the wave heating rate $Q_{\rm A} (s)$ is balanced by radiative and conductive
losses:
\begin{equation}
B_0 \frac{d} {d s} \left( \frac{F} {B_0} \right) = Q_{\rm A} (s) - n_e^2 \Lambda (T) ,
\label{eq:loop}
\end{equation}
where $F(s)$ is the sum of conductive and enthalpy fluxes:
\begin{equation}
F(s) =  - \kappa(T) \frac{dT}{ds} + 2.5 \mu_0 B_0 c_1 k_B (T - T_0) .  \label{eq:Fs}
\end{equation}
Here $\mu_0 \equiv n_p v / B_0$ is the proton flux per unit magnetic flux, which is constant
along the loop. The radiative loss function
$\Lambda (T)$ is taken from CHIANTI version 8 \citep[][]{Dere1997, DelZanna2015a},
assuming coronal abundances \citep[][]{Schmelz2012}. In the corona the thermal conductivity
$\kappa (T)$ is dominated by the electrons \citep[][]{Spitzer1962}, but in the TR there is an
important contribution from ambipolar diffusion, i.e., the upward diffusion of neutrals and
downward diffusion of ions and electrons \citep[][]{Fontenla1990, Fontenla1991}.
In this paper we use the following expression for the total conductivity:
\begin{equation}
\kappa (T) = \kappa_0 T^{5/2} + \frac{\kappa_1}{T} \left[ 1 + \left( \frac{T}{10^4 ~ \rm K}
\right)^{-5} \right] ,
\end{equation}
where $\kappa_0 = 10^{-6}$ $\rm erg ~ cm^{-1} ~ s^{-1} ~ K^{-7/2}$ is the Spitzer conductivity,
and the second term with $\kappa_1 = 5 \times 10^9$ $\rm erg ~ cm^{-1} ~ s^{-1}$ is an
approximation for the contribution from ambipolar diffusion \citep[based on Figure~7 in][]
{Fontenla1991}. The heating rate is assumed to be of the form given in equation (\ref{eq:QA}).
Then equation (\ref{eq:loop}) can be integrated as follows:
\begin{equation}
\frac{F(s)} {B_0 (s)} = \frac{F(0)} {B_0 (0)}  + c_0 \int_0^{s} [ B_0 (s) ]^{n-1} ds
- \int_0^{s} \frac{n_e^2 \Lambda (T)} {B_0 (s)} ds .  \label{eq:FB}
\end{equation}
Using $s = L_{\rm c}$, we obtain an expression for $c_0$ in terms of two integrals along the
coronal loop and values of the fluxes $F(0)$ and $F(L_{\rm c})$ at the two endpoints. Given
an estimate for the coronal temperature $T(s)$ and pressure $p_{\rm cor}$, both integrals can
be readily computed.
The endpoint fluxes are computed as described in the Appendix of \citet[][] {Schrijver2005}.
Therefore, equation (\ref{eq:FB}) with $s = L_{\rm c}$ yields the value of $c_0$, which can be
used to derive the heating rate $Q_{\rm A} (s)$ and total energy flux $F(s)$. Similarly, equation
(\ref{eq:Fs}) can be integrated to yield
\begin{equation}
\int_0^{s} F(s) ds = - \int_{T(0)}^{T(s)} \kappa (T) dT + 
2.5 \mu_0 c_1 k_B \int_0^{s} B_0 (s) [T(s) - T_0] ds ,
\label{eq:Fint}
\end{equation}
and using $s = L_{\rm c}$ we find an expression for mass flow parameter $\mu_0$.
Here we use the fact that the temperatures at the two endpoints are equal, $T (0) = T(L_{\rm c})
= 2 T_0$, so the integral over the conductive flux vanishes when integrating along the entire loop.
Finally, we obtain the quantity $\int_{T(0)}^{T(s)} \kappa (T) dT$ as a function of $s$ from
equation (\ref{eq:Fint}), which can be inverted to obtain a new estimate for the temperature
$T(s)$. We repeat this process until the relative changes in temperature are less than $0.1\%$
everywhere along the loop.


\clearpage

\begin{thebibliography}{}

\bibitem[Abramenko et al.(2011)]{Abramenko2011}
Abramenko, V.I., Carbone, V., Yurchyshyn, V., et al. 2011, \apj,
743, 133

\bibitem[Alfv\'{e}n(1947)]{Alfven1947}
Alfv\'{e}n, H., 1947, \mnras, 107, 211

\bibitem[Antolin \& Rouppe van der Voort(2012)]{Antolin2012}
Antolin, P., \& Rouppe van der Voort, L. 2012, \apj, 745, 152

\bibitem[Antolin \& Shibata(2010)]{Antolin2010}
Antolin, P., \& Shibata, K. 2010, \apj, 712, 494

\bibitem[Antolin et al.(2008)]{Antolin2008}
Antolin, P., Shibata, K., Kudoh, T., Shiota, D., \& Brooks, D. 2008, \apj, 688, 669

\bibitem[Antolin et al.(2015)]{Antolin2015}
Antolin, P., Vissers, G., Pereira, T. M. D., Rouppe van der Voort, L., \& Scullion, E.
2015, \apj, 806, 81

\bibitem[Arregui(2015)]{Arregui2015}
Arregui, I. 2015, Phil. Trans. R. Soc. A 373,
20140261, http://dx/doi.org/10.1098/rsta.2014.0261

\bibitem[Aschwanden(2005)]{Aschwanden2005}
Aschwanden, M. 2005, Physics of the Solar Corona (Springer: Berlin)

\bibitem[Asgari-Targhi \& van Ballegooijen(2012)]{Asgari2012}
Asgari-Targhi, M., \& van Ballegooijen, A.A. 2012, \apj, 746, 81 (paper~II)

\bibitem[Asgari-Targhi et al.(2013)]{Asgari2013} 
Asgari-Targhi, M., van Ballegooijen, A.A., Cranmer, S.R., \&
DeLuca, E.E. 2013, \apj, 773, 111 (paper~III)

\bibitem[Asgari-Targhi et al.(2014)]{Asgari2014}
Asgari-Targhi, M., van Ballegooijen, \& Imada, S. 2014, \apj, 786, 28 (paper~IV)

\bibitem[Asgari-Targhi et al.(2015)]{Asgari2015}
Asgari-Targhi, M., Schmelz, J. T., Imada, S., Pathak, S., \& Christian, G. M.
2015, \apj, 807, 146

\bibitem[Bale et al.(2005)]{Bale2005}
Bale, S. D., Kellogg, P. J., Mozer, F. S., Horbury, T. S., \& Reme, H.
2005, Phys. Rev. Letters, 94, 215002

\bibitem[Beckers(1972)]{Beckers1972}
Beckers, J. M. 1972, ARA\&A, 10, 73

\bibitem[Belcher(1971)]{Belcher1971}
Belcher, J.W. 1971, \apj, 168, 509

\bibitem[Berger \& Title(1996)]{Berger1996}
Berger, T. E., \& Title, A. M. 1996, \apj, 463, 365

\bibitem[Berger \& Title(2001)]{Berger2001}
Berger, T. E., \& Title, A. M. 2001, \apj, 553, 449

\bibitem[Berger et al.(1998)]{Berger1998}
Berger, T. E., L\"{o}fdahl, M. G., Shine, R. A., \& Title, A. M. 1998, \apj, 506, 439

\bibitem[Berger et al.(2004)]{Berger2004}
Berger, T. E., Rouppe van der Voort, L. H. M., L\"{o}fdahl, M. G., Carlsson, M.,
Fossum, A., Hansteen, V. H., Marthinussen, E., Title, A., \& Scharmer, G.
2004, \aap, 428, 613

\bibitem[Berger(1991)]{Berger1991}
Berger, M. A. 1991, \aap, 252, 369

\bibitem[Berger(1993)]{Berger1993}
Berger, M. A. 1993, Phys. Rev. Lett., 70, 705

\bibitem[Berger \& Asgari-Targhi(2009)]{Berger2009}
Berger, M. A., \& Asgari-Targhi, M. 2009, \apj, 705, 347

\bibitem[Berger et al.(2015)]{Berger2015}
Berger, M. A., Asgari-Targhi, M., \& DeLuca, E. E. 2015,
J. Plasma Phys., 81, 395810404

\bibitem[Bian et al.(2016a)]{Bian2016a}
Bian, N. H., Kontar, E. P., \& Emslie, A. G. 2016, \apj, 824, 78

\bibitem[Bian et al.(2016b)]{Bian2016b}
Bian, N. H., Watters, J., M., Kontar, E. P., \& Emslie, A. G. 2016, \apj, 833, 76

\bibitem[Bingert \& Peter(2011)]{Bingert2011}
Bingert, S., \& Peter, H. 2011, \aap, 530, A112

\bibitem[Borovsky(2012)]{Borovsky2012}
Borovsky, J. E. 2012, \jgr, 117, A05104

\bibitem[Borrero \& Ichimoto(2011)]{Borrero2011}
Borrero, J. M., \& Ichimoto, K. 2011, Living Rev. Solar Phys., 8, 4,
http://www.livingreviews.org/lrsp-2011-4

\bibitem[Bourdin et al.(2015)]{Bourdin2015}
Bourdin, Ph.-A., Bingert, S., \& Peter, H. 2015, \aap, 580, A72

\bibitem[Brooks et al.(2013)]{Brooks2013}
Brooks, D. H., Warren, H. P., Ugarte-Urra, I., \& Winebarger, A. R. 2013, \apjl, 772, L16

\bibitem[Brooks \& Warren(2016)]{Brooks2016}
Brooks, D. H., \& Warren, H. P. 2016, \apj, 820, 63

\bibitem[Brosius et al.(1996)]{Brosius1996}
Brosius, J. W., Davila, J. M., Thomas, R. J., et al. 1996, \apjs, 106, 143

\bibitem[Bruls \& Solanki(1995)]{Bruls1995}
Bruls, J. H. M. J., \& Solanki, S. K. 1995, \aap, 293, 240

\bibitem[Buehler et al.(2015)]{Buehler2015}
Buehler, D., Lagg, A., Solanki, S. K., \& van Noort, M. 2015, \aap, 576, A27

\bibitem[Cargill(1994)]{Cargill1994}
Cargill, P. 1994, \apj, 422, 381

\bibitem[Cargill et al.(2015)]{Cargill2015}
Cargill, P. J., Warren, H. P., \& Bradshaw, S. J. 2015, Phil. Trans. R. Soc. A 373,
20140260, http://dx/doi.org/10.1098/rsta.2014.0260

\bibitem[Cargill \& Klimchuk(1997)]{Cargill2004}
Cargill, P., \& Klimchuk, J. A. 1997, \apj, 478, 799

\bibitem[Cargill \& Klimchuk(2004)]{Cargill1997}
Cargill, P., \& Klimchuk, J. A. 2004, \apj, 605, 911

\bibitem[Chandran et al.(2011)]{Chandran2011}
Chandran, B. D. G., Dennis, T. J., Quataert, E., \& Bale, S. D. 2011, \apj, 743, 197

\bibitem[Chitta et al.(2012)]{Chitta2012}
Chitta, L.P., van Ballegooijen, A.A., Rouppe van der Voort, L.,
DeLuca, E. E., \& Kariyappa, R. 2012, \apj, 752, 48

\bibitem[Cho et al.(2002)]{Cho2002}
Cho, J., Lazarian, A., \& Vishniac, E. T. 2002, \apj, 564, 291

\bibitem[Cohen et al.(1950)]{Cohen1950}
Cohen, R. S., Spitzer, L., Jr. \& Routly, P. M. 1950, Phys. Rev., 80, 230

\bibitem[Coleman(1968)]{Coleman1968}
Coleman, P. J., Jr. 1968, \apj, 153, 371

\bibitem[Cranmer et al.(2015)]{Cranmer2015}
Cranmer, S.R., Asgari-Targhi, M., Miralles, M. P., Raymond, J. C.,
Strachan, L., Tian, H., \& Woolsey, L. N. 2015,
Phil. Trans. Royal Soc. A, 373, 20140148

\bibitem[Cranmer \& Woolsey(2015)]{Cranmer2015b}
Cranmer, S. R.,  \& Woolsey, L. N. 2015, \apj, 812, 71

\bibitem[Cranmer et al.(2007)]{Cranmer2007}
Cranmer, S. R., van Ballegooijen, A. A., \& Edgar, R. J. 2007, \apjs, 171, 520

\bibitem[Dahlburg et al.(2016)]{Dahlburg2016}
Dahlburg, R. B., Einaudi, G., Taylor, B. D., Ugarte-Urra, I., Warren, H. P.,
Rappazzo, A. F., \& Velli, M. 2016, \apj, 817, 47

\bibitem[Defouw(1976)]{Defouw1976}
Defouw, R. J. 1976, \apj, 209, 266

\bibitem[De Groof \& Goossens(2002)]{DeGroof2002}
De Groof, A., \& Goossens, M. 2002, \aap, 386, 691

\bibitem[DelZanna et al.(2015a)]{DelZanna2015a}
DelZanna, G., Dere, K. P., Young, P. R., Landi, E., \& Mason, H. 2015a, \aap, 582, A56

\bibitem[DelZanna et al.(2015b)]{DelZanna2015b}
DelZanna, G., Tripathi, D., Mason, H., Subramanian, S., \& O'Dwyer, B. 2015b, \aap, 573, A104

\bibitem[De Moortel \& Browning(2015)]{DeMoortel2015}
De Moortel, I., \& Browning, Ph. 2015, Phil. Trans. R. Soc. A 373,
20140269, http://dx/doi.org/10.1098/rsta.2014.0269

\bibitem[De Pontieu et al.(2007)]{DePontieu2007}
De Pontieu, B., McIntosh, S. W., Carlsson, M., et al. 2007, Sci, 318, 1574

\bibitem[Dere et al.(1997)]{Dere1997}
Dere, et al. 1997, \aaps, 125, 149

\bibitem[DeVore et al.(1985)]{DeVore1985}
DeVore, C. R., Sheeley, N. R., Jr., Boris, J. P., Young, T. R., Jr., \&
Harvey, K., 1985, \solphys, 102, 41

\bibitem[Dmitruk \& Matthaeus(2003)]{Dmitruk2003}
Dmitruk, P., \& Matthaeus, W. H. 2003, \apj, 597, 1097

\bibitem[Doschek(2012)]{Doschek2012}
Doschek, G. A. 2012, \apj, 754, 153 

\bibitem[Doschek et al.(2007)]{Doschek2007}
Doschek, G. A., Mariska, J. T., Warren, H. P., Brown, C. M., Culhane, J. L. ,
Watanabe, T., Young, P. R., \& Mason, H. E. 2007, \apjl, 667, L109 

\bibitem[Elsasser(1950)]{Elsasser1950}
Elsasser, W. M. 1950, Phys. Rev., 79, 183

\bibitem[Feldman et al.(1997)]{Feldman1997}
Feldman, U., Doschek, G. A., \& Mariska, J. T. 1979, \apj, 229, 369

\bibitem[Fischer et al.(2009)]{Fischer2009}
Fischer, C. E., de Wijn, A. G., Centeno, R., Lites, B.. W., \& Keller, C. U. 2009,
\aap, 504, 583

\bibitem[Fletcher \& de Pontieu(1999)]{Fletcher1999}
Fletcher, L., \& de Pontieu, B. 1999, \apjl, 520, L135

\bibitem[Fludra et al.(2017)]{Fludra2017}
Fludra, A., Hornsey, C., \& Nakariakov, V. M. 2017, \apj, 834, 100

\bibitem[Fontenla et al.(1990)]{Fontenla1990}
Fontenla, J. M., Avrett, E. H., \& Loeser, R. 1990, \apj, 355, 700

\bibitem[Fontenla et al.(1991)]{Fontenla1991}
Fontenla, J. M., Avrett, E. H., \& Loeser, R. 1991, \apj, 377, 712

\bibitem[Galsgaard \& Nordlund(1996)]{Galsgaard1996}
Galsgaard, K., \& Nordlund, {\AA} 1996, \jgr, 101, 13445

\bibitem[Goldreich \& Sridhar(1995)]{Goldreich1995}
Goldreich, P., \& Sridhar, S. 1995, \apj, 438, 763

\bibitem[Goossens et al.(2012)]{Goossens2012}
Goossens, M., Andries, J., Soler, R., et al. 2012, \apj, 753, 111

\bibitem[Goossens et al.(2011)]{Goossens2011}
Goossens, M., Erd\'{e}lyi, R., Ruderman, M.S. 2011, \ssr, 158, 289

\bibitem[Goossens et al.(2013)]{Goossens2013}
Goossens, M., Van Doorsselaere, T., Soler, R., \& Verth, G. 2013, \apj, 768, 191

\bibitem[Gudiksen \& Nordlund(2005)]{Gudiksen2005}
Gudiksen, B. V., \& Nordlund, {\AA} 2005, \apj, 618, 1020

\bibitem[Hagenaar et al.(1999)]{Hagenaar1999}
Hagenaar, H. J., Schrijver, C. J., Title, A. M., \&b Shine, R., A. 1999, \apj, 511, 932

\bibitem[Hara \& Ichimoto(1999)]{Hara1999}
Hara, H., \& Ichimoto, K. 1999, \apj, 513: 969

\bibitem[Hara et al.(2008)]{Hara2008}
Hara, H., et al. 2008, \apj, 678, L67 

\bibitem[Heyvaerts \& Priest(1983)]{Heyvaerts1983}
Heyvaerts, J., \& Priest, E.R. 1983, \aap, 117, 220

\bibitem[Hollweg(1978)]{Hollweg1978}
Hollweg, J. V. 1978, \solphys, 56, 305

\bibitem[Hollweg(1986)]{Hollweg1986}
Hollweg, J. V. 1986, \jgr, 91, 4111

\bibitem[Hollweg et al.(1982)]{Hollweg1982}
Hollweg, J. V., Jackson, S., \& Galloway, D. 1982, \solphys, 75, 35

\bibitem[Ichimoto et al.(1995)]{Ichimoto1995}
Ichimoto, K., Hara, H., Takeda, A., Kumagai, K., Sakurai, T., Shimizu, T., \&
Hudson, H. S. 1995, \apj, 445, 978

\bibitem[Iroshnikov(1963)]{Iroshnikov1963} 
Iroshnikov, P. S. 1963, Astron. Zh., 40, 742 (English translation in
Sov. Astron. 7, 566 [1964])

\bibitem[Janse et al.(2010)]{Janse2010}
Janse, A. M., Low, B. C., \& Parker, E. N. 2010, Phys. Plasmas, 17, 092901

\bibitem[Kano \& Tsuneta(1996)]{Kano1996}
Kano, R., \& Tsuneta, S. 1996, \pasj, 48, 535

\bibitem[Klimchuk(2000)]{Klimchuk2000}
Klimchuk, J. A. 2000, \solphys, 193, 53

\bibitem[Klimchuk(2006)]{Klimchuk2006}
Klimchuk, J. A. 2006, \solphys, 234, 41

\bibitem[Klimchuk(2010)]{Klimchuk2010}
Klimchuk, J. A. 2010, in ASP Conf. Ser. 415, Proc. of Second Hinode Science Meeting,
ed. B. Lites et al. (San Francisco: ASP), 221

\bibitem[Komm et al.(1995)]{Komm1995}
Komm, R. W., Howard, R. F., \& Harvey, J. W. 1995, \solphys, 158, 213

\bibitem[Kontar et al.(2014)]{Kontar2014}
Kontar, E. P., Bian, N. H., Emslie, A. G., \& Vilmer, N. 2014, \apj, 780, 176

\bibitem[Kraichnan(1965)]{Kraichnan1965}
Kraichnan, R. H. 1965, Phys. Fluids, 8, 1385

\bibitem[Kudoh \& Shibata(1999)]{Kudoh1999}
Kudoh, T., \& Shibata, K. 1999, \apj, 514, 493

\bibitem[Lionello et al.(2013)]{Lionello2013}
Lionello, R., Winebarger, A. R., Mok, Y., Linker, J. A., \& Miki\'{c}, Z. 2013, \apj, 773, 134

\bibitem[L\'{o}pez-Fuentes et al.(2006)]{LopezFuentes2006}
L\'{o}pez-Fuentes, M., Klimchuk, J. A., \& D\'{e}moulin, P. 2006, \apj, 639, 459

\bibitem[L\'{o}pez-Fuentes et al.(2008)]{LopezFuentes2008}
L\'{o}pez-Fuentes, M., D\'{e}moulin, P., \& Klimchuk, J. A. 2008, \apj, 673, 586

\bibitem[L\'{o}pez-Fuentes \& Klimchuk(2015)]{LopezFuentes2015}
L\'{o}pez-Fuentes, M., \& Klimchuk, J. A. 2015, \apj, 828, id.86

\bibitem[L\'{o}pez-Fuentes \& Klimchuk(2016)]{LopezFuentes2016}
L\'{o}pez-Fuentes, M., \& Klimchuk, J. A. 2016, \apj, 799, id.128

\bibitem[Manso Sainz et al.(2011)]{MansoSainz2011}
Manso Sainz, R., Mart\'{i}nez Gonz\'{a}lez, M. J., \& Asensio Ramos, E. 2011, \aap, 531, L9

\bibitem[Maron \& Goldreich(2001)]{Maron2001}
Maron, J., \& Goldreich, P. 2001, \apj, 554, 1175

\bibitem[Martens et al.(2000)]{Martens2000}
Martens, P. C. H., Kankelborg, C. C., \& Berger, T. E. 2000, \apj, 537, 471

\bibitem[Mart\'{i}nez Pillet et al.(1997)]{MartinezPillet1997}
Mart\'{i}nez Pillet, V., Lites, B. W., \& Skumanich, A. 1997, \apj, 474, 810

\bibitem[Matthaeus et al.(1990)]{Matthaeus1990}
Matthaeus, W. H., Goldstein, M. L., \& Roberts, D. A. 1990, JGR, 95, 20673

\bibitem[Matthaeus et al.(1999)]{Matthaeus1999}
Matthaeus, W. H., Zank, G. P., Oughton, S. Mullan, D. J., \& Dmitruk, P. 1999, \apj, 523, L93

\bibitem[Matsumoto \& Shibata(2010)]{Matsumoto2010}
Matsumoto, T., \& Shibata, K. 2010, \apj, 710, 1857

\bibitem[McIntosh et al.(2008)]{McIntosh2008}
McIntosh, S. W., De Pontieu, B., \& Tarbell, T. D. 2008, \apj, 673, L219

\bibitem[McIntosh et al.(2011)]{McIntosh2011}
McIntosh, S. W., De Pontieu, B., Carlsson, M., Hansteen, V., Boerner, P., \&
Goossens, M. 2011, Nature, 475, 477

\bibitem[Mercier \& Trottet(1997)]{Mercier1997}
Mercier, C., \& Trottet, G. 1997, \apj, 474, L65

\bibitem[Miki\'{c} et al.(1989)]{Mikic1989}
Miki\'{c}, Z., Schnack, D. D., \& van Hoven, G. 1989, \apj, 338, 1148

\bibitem[Miki\'{c} et al.(2013)]{Mikic2013}
Miki\'{c}, Z., Lionello, R., Mok, Y., Linker, J. A., \& Winebarger, A. R. 2013, \apj, 773, 94

\bibitem[Moriyasu et al.(2004)]{Moriyasu2004}
Moriyasu, S., Kudoh, T., Yokoyama, T., \& Shibata, K. 2004, \apj, 601, L107

\bibitem[Morton et al.(2015)]{Morton2015}
Morton, R. J., Tomczyk, S., \& Pinto, R. 2015, Nat. Comm., DOI: 10.1038/ncomms8813

\bibitem[Muller et al.(1994)]{Muller1994}
Muller, R., Roudier, T., Vigneau, J. \& Auffret, H. 1994, \aap, 283, 232

\bibitem[Mumford et al.(2015)]{Mumford2015}
Mumford, S. J., Fedun, V., \& Erd\'{e}lyi, R. 2015, \apj, 799, 6

\bibitem[Murawski et al.(2015)]{Murawski2015}
Murawski, K., Solovev, A. Musielak, K. et al. 2015, \aap, 577, A126

\bibitem[Nagata et al.(2008)]{Nagata2008}
Nagata, S., Tsuneta, S., Suematsu, Y., et al. 2008, \apj, 677, L145

\bibitem[Nisenson et al.(2003)]{Nisenson2003}
Nisenson, P., van Ballegooijen, A. A., de Wijn, A. G., \& S\"{u}tterlin, P. 2003, \apj, 587, 458

\bibitem[Ortiz et al.(2014)]{Ortiz2014}
Ortiz, A., Bellot Rubio, L. R., Hansteen, V. H., De La Cruz Rodr\'{i}guez, J., \&
Rouppe Van Der Voort, L. 2014, \apj, 781, 126

\bibitem[Oughton et al.(2001)]{Oughton2001}
Oughton, S., Matthaeus, W. H., Dmitruk, P., Milano, L. J.,
Zank, G. P., \& Mullan, D. J. 2001, \apj, 551, 565

\bibitem[Parnell \& De Moortel(2012)]{Parnell2012}
Parnell, C. E., \& De Moortel, I. 2012,  Phil. Trans. R. Soc. A 370, 3217-3240

\bibitem[Parker(1972)]{Parker1972}
Parker, E. N. 1972, \apj, 174, 499

\bibitem[Parker(1978)]{Parker1978}
Parker, E. N. 1978, \apj, 221, 368

\bibitem[Parker(1983)]{Parker1983}
Parker, E. N. 1983, \apj, 264, 642

\bibitem[Parker(1988)]{Parker1988}
Parker, E. N. 1988, \apj, 330, 474

\bibitem[Pascoe et al.(2012)]{Pascoe2012}
Pascoe, D. J., Hood, A. W., de Moortel, I., \& Wright, A. N. 2012,
\aap, 539, A37

\bibitem[Pascoe et al.(2011)]{Pascoe2011}
Pascoe, D. J., Wright, A. N., \& de Moortel, I. 2011, \apj, 731, 73

\bibitem[Patsourakos \& Klimchuk(2009)]{Patsourakos2009}
Patsourakos, S., \& Klimchuk, J. A. 2009, \apj, 696, 760

\bibitem[Perez \& Chandran(2013)]{Perez2013}
Perez, J. C., \& Chandran, B. D. G. 2013, \apj, 776, 124

\bibitem[Peter(2013)]{Peter2013}
Peter, H. 2013, \solphys, 288, 531

\bibitem[Poedts et al.(1990)]{Poedts1990}
Poedts, S., Goossens, M., \& Kerner, W. 1990, \apj, 360, 279

\bibitem[Pontin \& Hornig(2015)]{Pontin2015}
Pontin, D. I., \& Hornig, G. 2015, \apj, 805, 47

\bibitem[Pontin et al.(2017)]{Pontin2017}
Pontin, D. I., Janvier, M., Tiwari, S. K., Galsgarrd, K., Winebarger, A. R.,
\& Cirtain, J. W. 2017, \apj, 837, 108

\bibitem[Priest et al.(2002)]{Priest2002}
Priest, E. R., Heyvaerts, J. F., \& Title, A. M. 2002, \apj, 576, 533

\bibitem[Rappazzo et al.(2007)]{Rappazzo2007}
Rappazzo, A., F., Velli, M., Einaudi, G., \& Dahlburg, R. B. 2007, \apj, 657, L47

\bibitem[Rappazzo et al.(2008)]{Rappazzo2008}
Rappazzo, A., F., Velli, M., Einaudi, G., \& Dahlburg, R. B. 2008, \apj, 677, 1348

\bibitem[Rappazzo et al.(2013)]{Rappazzo2013}
Rappazzo, A., F., Velli, M., \& Einaudi, G. 2013, \apj, 771, 76

\bibitem[Richie et al.(2016)]{Ritchie2016}
Ritchie, M. L., Wilmot-Smith, A. L., \& Hornig, G. 2016, \apj, 824, id.19

\bibitem[Rosner et al.(1978)]{Rosner1978}
Rosner, R., Tucker, W. H., \& Vaiana, G. S. 1978, \apj, 220, 643

\bibitem[Schekochihin et al.(2009)]{Schekochihin2009}
Schekochihin, A. A., Cowley, S. C., Dorland, W., Hammett, G. W., Howes,
G. G., Quatert, E., \& Tatsuno, T. 2009, \apjs, 182, 310

\bibitem[Schmelz et al.(2012)]{Schmelz2012}
Schmelz, J. T., Reames, D. V., von Steiger, R., \& Basu, S. 2012, \apj, 755, 33

\bibitem[Schrijver \& Martin(1990)]{Schrijver1990}
Schrijver, C. J., \& Martin, S. F. 1990, \solphys, 129, 95

\bibitem[Schrijver et al.(1996)]{Schrijver1996}
Schrijver et al. 1996, \apj, 468, 921

\bibitem[Schrijver et al.(1999)]{Schrijver1999}
Schrijver et al. 1999, \solphys, 187, 261

\bibitem[Schrijver \& van Ballegooijen(2005)]{Schrijver2005}
Schrijver, C. J., \& van Ballegooijen, A. A. 2005, \apj, 630, 552

\bibitem[Schrijver \& Zwaan(2000)]{Schrijver2000}
Schrijver, C. J., \& Zwaan, C. 2000, {\it Solar and Stellar Magnetic Activity} (Cambridge
University Press: Cambridge, UK)

\bibitem[Scullion et al.(2014)]{Scullion2014}
Scullion, E., Rouppe van der Voort, L., Wedemeyer, S., \& Antolin, P. 2014,
\apj, 797, 36

\bibitem[Shebalin et al.(1983)]{Shebalin1983}
Shebalin, J. V., Matthaeus, W. H., \& Montgomery, D. 1983, J. Plasma Phys., 29, 525

\bibitem[Sheeley(1981)]{Sheeley1981}
Sheeley, N. R., Jr. 1981, The overall structure and evolution of active regions,
in {\it Solar Active Regions}, ed. Frank Q. Orrall (Colorado Associated University Press:
Boulder, Colorado), p. 17

\bibitem[Solanki(1993)]{Solanki1993}
Solanki, S. 1993, Space Sci Rev, 63, 1

\bibitem[Spitzer \& H\"{a}rm(1953)]{Spitzer1953}
Spitzer, L., Jr., \& H\"{a}rm, R. 1953, Phys. Rev., 89, 977

\bibitem[Spitzer(1962)]{Spitzer1962}
Spitzer, L., Jr. 1962, {\it Physics of Fully Ionized Gases} (New York: John Wiley and Sons, Inc)

\bibitem[Spruit(1976)]{Spruit1976}
Spruit, H. C. 1976, \solphys, 61, 363

\bibitem[Spruit(1979)]{Spruit1979}
Spruit, H. C. 1979, \solphys, 50, 269

\bibitem[Stenflo(1973)]{Stenflo1973}
Stenflo, J. O. 1973, \solphys, 32, 41

\bibitem[Sterling(2000)]{Sterling2000}
Sterling, A. C. 2000, \solphys, 196, 79

\bibitem[Strauss(1976)]{Strauss1976}
Strauss, H.R. 1976, Phys. Fluids, 19, 134

\bibitem[Strauss(1997)]{Strauss1997}
Strauss, H. R. 1997, J. Plasma Phys., 57, 83

\bibitem[Suzuki \& Inutsuka(2005)]{Suzuki2005}
Suzuki, T. K., \& Inutsuka, S.-I. 2005, \apj, 632, L49

\bibitem[Testa et al.(2016)]{Testa2016}
Testa, P, De Pontieu, B., \& Hansteen, V. 2016, \apj, 827, 99

\bibitem[Threlfall et al.(2013)]{Threlfall2013}
Threlfall, J., De Moortel, I., McIntosh, S. W., \& Bethge, C. 2013,
\aap, 556, A124

\bibitem[Tian et al.(2014)]{Tian2014}
Tian, H., DeLuca, E. E., Cranmer, S. R., et al. 2014, Science, 346,
1255711

\bibitem[Tian et al.(2011)]{Tian2011}
Tian, H., McIntosh, S. W., Habbal, S. R., He, J. 2011, \apj, 736, 130

\bibitem[Tian et al.(2012a)]{Tian2012a}
Tian, H., McIntosh, S. W. , Xia, L. , He, J. , \& Wang, X.  2012, \apj, 748, 106

\bibitem[Tian et al.(2012b)]{Tian2012b}
Tian, H., McIntosh, S. W. , Wang, T., Ofman, L., De Pontieu, B.,
Innes, D. E. , \& Peter, H. 2012, \apj, 759, 144

\bibitem[Tomczyk \& McIntosh(2009)]{Tomczyk2009}
Tomczyk, S., \& McIntosh, S. W. 2009, \apj, 697, 1384

\bibitem[Tomczyk et al.(2007)]{Tomczyk2007}
Tomczyk, S., McIntosh, S. W., Keil, S. L., et al. 2007, Sci, 317, 1192

\bibitem[Tripathi et al.(2011)]{Tripathi2011}
Tripathi, D., Klimchuk, J. A., \& Mason, H. E. 2011, \apj, 740, 111

\bibitem[Tripathi et al.(2009)]{Tripathi2009} 
Tripathi, D., Mason, H. E., Dwivedi, B. N., Del Zanna, G., \& Young,
P. R. 2009, \apj, 694, 1256

\bibitem[Utz et al.(2010)]{Utz2010}
Utz, D., Hanslmeier, R., Muller, R., et al. 2010, \aap, 511, A39

\bibitem[van Ballegooijen(1986)]{vanB1986}
van Ballegooijen, A. A. 1986, \apj, 311, 1001

\bibitem[van Ballegooijen(1988)]{vanB1988}
van Ballegooijen, A. A. 1988, in {\it Solar and Stellar Coronal Structure and Dynamics}, 
Proc. of the Ninth Sacramento Peak Summer Symposium, ed. R. C. Altrock (National Solar
Observatory: Sunspot, NM 88349), p. 115

\bibitem[van Ballegooijen \& Hasan(2003)]{vanB2003}
van Ballegooijen, A. A., \& Hasan, S. S. 2003, in {\it Current Theoretical Models and
High Resolution Solar Observations: Preparing for ATST}, eds. A. A. Pevtsov, \&
H. Uitenbroek, ASP Conf. Series, Vol. 286, p.155

\bibitem[van Ballegooijen et al.(1998)]{vanB1998}
van Ballegooijen, A. A., Nisenson, P., Noyes, R. W., L\"{o}fdahl, M. G.,
Stein, R. F., Nordlund, {\AA}., \& Krishnakumar, V. 1998, \apj, 509, 435

\bibitem[van Ballegooijen et al.(2014)]{vanB2014}
van Ballegooijen, A. A., Asgari-Targhi, M., \& Berger, M. A. 2014, \apj, 787, 87

\bibitem[van Ballegooijen et al.(2011)]{vanB2011}
van Ballegooijen, A. A., Asgari-Targhi, M., Cranmer, S. R., \&
DeLuca, E. E. 2011, \apj, 736, article 3 (paper~I)

\bibitem[van Ballegooijen \& Asgari-Targhi(2016)]{vanB2016}
van Ballegooijen, A. A., \& Asgari-Targhi, M. 2016, \apj, 821, 106

\bibitem[van Ballegooijen \& Asgari-Targhi(2017)]{vanB2017}
van Ballegooijen, A. A., \& Asgari-Targhi, M. 2017, \apj, 835, 10 (paper~V)

\bibitem[Verdini \& Velli(2007)]{Verdini2007}
Verdini, A., \& Velli, M. 2007, \apj, 662, 669

\bibitem[Viall \& Klimchuk(2011)]{Viall2011}
Viall, N. M., \& Klimchuk, J. A. 2011, \apj, 738, 24

\bibitem[Viall \& Klimchuk(2012)]{Viall2012}
Viall, N. M., \& Klimchuk, J. A. 2012, \apj, 753, 35

\bibitem[Vigeesh et al.(2012)]{Vigeesh2012}
Vigeesh, G., Fedun, V., Hasan, S. S., \& Erd\'{e}lyi, R. 2012, \apj, 755, 18

\bibitem[Wentzel(1974)]{Wentzel1974}
Wentzel, D. G. 1974, \solphys, 39, 129

\bibitem[Wang(1988)]{Wang1988}
Wang, H. 1988, \solphys, 116, 1

\bibitem[Warren et al.(2008)]{Warren2008}
Warren, H. P., Winebarger, A. R., Mariska, J. T., Doschek, G. A., \&
Hara, H. 2008, \apj, 677, 1395

\bibitem[Warren et al.(2011)]{Warren2011}
Warren, H. P., Brooks, D. H., \& Winebarger, A. R. 2011, \apj, 734, 90

\bibitem[Warren et al.(2012)]{Warren2012}
Warren, H. P., Winebarger, A. R., \& Brooks, D. H. 2012, \apj, 759, 141

\bibitem[Wedemeyer-B\"{o}hm et al.(2012)]{Wedemeyer2012}
Wedemeyer-B\"{o}hm, S., Scullion, E. Steiner, O., et al. 2012, \nat, 486, 505

\bibitem[Winebarger et al.(2011)]{Winebarger2011}
Winebarger, A. R., Schmelz, J. T., Warren, H. P., Saar, S. H., \& Kashyap, V. L.
2011, \apj, 740, 2

\bibitem[Wilmot-Smith et al.(2009)]{Wilmot-Smith2009}
Wilmot-Smith, A. L., Hornig, G., \& Pontin, D. I. 2009, \apj, 704, 1288

\bibitem[Wilmot-Smith(2015)]{Wilmot-Smith2015}
Wilmot-Smith, A. L. 2015, Phil. Trans. R. Soc. A 373,
20140265, http://dx/doi.org/10.1098/rsta.2014.0265

\bibitem[Yelles Chaouche et al.(2009)]{YellesChaouche2009}
Yelles Chaouche, L., Solanki, S. K., \& Sch\"{u}ssler, M. 2009, \aap, 504, 595

\bibitem[Young et al.(2007)]{Young2007}
Young, P. R., Del Zanna, G., Mason, H. E., et al. 2007, PASJ, 59, S857

\bibitem[Zhang et al.(1998)]{Zhang1998}
Zhang, J., White, S. M., \& Kundu, M. R. 1998, \apjl, 504, L127

\bibitem[Zirker(1993)]{Zirker1993}
Zirker, J. B. 1993, \solphys, 148, 43

\end{thebibliography}
\end{document}